\documentclass[useAMS,usenatbib,aps,floatdix,nofootinbib]{mn2e}

\usepackage{graphics}
\usepackage{amssymb,amsmath}
\usepackage{amsmath}
\usepackage{psfrag}
\usepackage{graphicx}
\newcommand{\ve}[1]{\mathbf{#1}}

\newcommand{\f}{\frac}

\newcommand{\be}{\begin{equation}}      
\newcommand{\ee}{\end{equation}}      
      
\newcommand{\bef}{\begin{figure}}      
\newcommand{\eef}{\end{figure}}      
\newcommand{\bea}{\begin{eqnarray}}    
\newcommand{\eea}{\end{eqnarray}}

\newcommand{\bx}{{\bf x}}

\def\spose#1{\hbox to 0pt{#1\hss}}
\def\ltapprox{\mathrel{\spose{\lower 3pt\hbox{$\mathchar"218$}}
\raise 2.0pt\hbox{$\mathchar"13C$}}}
\def\gtapprox{\mathrel{\spose{\lower 3pt\hbox{$\mathchar"218$}}
\raise 2.0pt\hbox{$\mathchar"13E$}}}
\def\inapprox{\mathrel{\spose{\lower 3pt\hbox{$\mathchar"218$}}
\raise 2.0pt\hbox{$\mathchar"232$}}}

\def\bx{{\bf x}}  

\def\br{{\bf r}}

\def\bse{\begin{subequations}}
\def\ese{\end{subequations}}

\def\lsim{\raise 0.4ex\hbox{$<$}\kern -0.8em\lower 0.62ex\hbox{$\sim$}} 
\def\gsim{\raise 0.4ex\hbox{$>$}\kern -0.7em\lower 0.62ex\hbox{$\sim$}}

\def\f0N{f_0^{(N)}}
\def\bec{\begin{center}}
\def\eec{\end{center}}

\title[Energy ejection in cold spherical collapse]
{Energy ejection in the collapse of a cold spherical 
self-gravitating cloud}
\author[M. Joyce, B. Marcos and F. Sylos Labini]
{M. Joyce$^1$, B. Marcos${^{2,}}{^{3,}}{^4}$ and F. Sylos Labini${^{2,}}{^{3}}$\\
$^1$Laboratoire de Physique Nucl\'eaire et de Hautes Energies,
UMR 7585, Universit\'e Pierre et Marie Curie --- Paris 6, \\
75252 Paris Cedex 05, France\\
$^{2}$``E. Fermi'' Center, Via Panisperna 89 A, Compendio del
Viminale, I-00184 Rome, Italy\\
$^3$ISC-CNR, Via dei Taurini 19, I-00185 Rome, Italy\\
$^4$Laboratoire J.-A. Dieudonn\'e, UMR 6621, 
Universit\'e de Nice --- Sophia Antipolis,
Parc Valrose 06108 Nice Cedex 02, France}

\begin{document}

\date{\today}

\maketitle

\begin{abstract}
When an open system of classical point particles interacting by 
Newtonian gravity collapses and relaxes violently, an arbitrary amount 
of energy may in principle be carried
away by particles which escape to infinity. We investigate here, using
numerical simulations, how this released energy and other related
quantities (notably the binding energy and size of the virialized
structure) depends on the initial conditions, for the one parameter 
family of starting configurations given by randomly distributing 
$N$ cold particles in a spherical volume.
Previous studies have 
established that the minimal size reached by the system scales
approximately as $N^{-1/3}$, a behaviour which follows trivially
when the growth of perturbations (which regularize the singularity
of the cold collapse in the $N\rightarrow \infty$ limit) 
are assumed to be unaffected by the boundaries.
Our study shows that the energy ejected grows approximately in
proportion to $N^{1/3}$, while the fraction of the initial mass 
ejected grows only
very slowly with $N$, approximately logarithmically, in the range of
$N$ simulated. We examine
in detail the mechanism of this mass and energy ejection, showing
explicitly that it arises from the interplay of the growth of
perturbations with the finite size of the system. A net lag of 
particles compared to their uniform spherical collapse trajectories
develops first at the boundaries and then propagates into the 
volume during the collapse. Particles in the outer shells are then 
ejected as they scatter through the time dependent potential of an 
already re-expanding central core. Using modified initial
configurations we explore the importance of fluctuations at
different scales, and discreteness (i.e. non-Vlasov) effects
in the dynamics.

\end{abstract}


\section{introduction}


%
That isolated systems of initially cold self-gravitating particles 
collapse and relax violently to produce virialized quasi-equilibrium
structures has been known for many decades --- essentially since
numerical simulation of such systems began (see, e.g.,
\cite{henon_1964}).
In cosmology it has emerged in the last decade
or so, through large numerical simulations, that a good approximate 
description of the structures formed can be given in terms of
``halos'', which are similar quasi-equilibrium structures formed from 
collapse of matter in finite regions, containing initially
cold (dark matter) particles (for a review, see e.g. \cite{halo}). 
Analytical understanding of the
strongly non-linear physics involved in the formation of these
quasi--equilibrium states is, despite the importance of the problem
and many attempts to solve it, extremely limited. For example, while
the simple density profiles which result from a class of uniform
initial conditions have been known since early numerical studies
(see, e.g., \cite{vanalbada_1982} and references therein), there is 
still no theory which can clearly 
explain them. Likewise, in 
cosmology, numerical evidence for the ``universality'' 
of certain simple halos profiles (different to those obtained
from quasi-uniform cold collapse) has been 
produced (\cite{navarro1, navarro2}), but understanding 
of the physical reason for the ubiquity of these profiles is 
still lacking. 

In attempts to gain greater insight into the physics leading
to such virialized states, with the goal notably of understanding
the dependence of their properties on initial conditions, one angle 
of approach, which is the one adopted here, is to study in detail 
the evolution from some limited class of initial conditions. We focus
here on what appears to us a neglected, and potentially important, 
aspect in the study of collapse and virialization of open 
self-gravitating systems: in the phase of (violent) relaxation
such systems may, in general, eject a fraction $f^p$ of their 
mass which carry away a finite amount of energy $K^p$ as
kinetic energy, i.e., some of the particles come out of the collapse 
with positive energy and may escape to 
infinity. The binding energy and characteristic size 
of the virialized state, in particular, are directly related
to these ejected quantities. Indeed, using (1) total energy 
and mass conservation, (2) the virial condition for the
bound particles, and (3) neglecting the potential energy 
between the bound and escaping particles, and that of 
the escaping particles themselves, it follows that 
\be
W^n=2(E_0-K^p)
\ee
where $W^n$ is the (negative) potential energy of the bound particles, 
and $E_0$ is the total initial energy. While for $E_0>0$ some energy 
must be ejected, for $E_0<0$ energy may, or may not, be ejected. 
In principle there is no upper bound on this ejected energy 
(as there is no lower bound on the gravitational potential). 
Indeed it is known that a self-gravitating system of
point particles is intrinsically unstable towards the ejection
of an infinite amount of energy --- the so-called ``gravo-thermal
catastrophe'' \citep{lyndenbell+wood_1968, binney}. This instability 
is understood, however, to be relevant 
in practice only on time scales which diverge with particle number
$N$. The term ``violent relaxation'', on the other hand, 
refers precisely to a physical relaxation process which takes 
places on much shorter scales, involve purely mean 
field (i.e. ``collisionless'') physics \citep{lyndenbell}. In the 
study we present in this
paper we will see that the energy ejected on these shorter time scales,
during violent relaxation, appears to be unbounded above for the class of 
initial conditions we study. Further we will investigate whether the 
ejection we observe in our simulations can be fully understood
as a mean field phenomenon.

The central considerations in this article --- energy ejection
and the validity of the mean field limit in cold collapse --- 
are of relevance in theoretical attempts to understand violent 
relaxation. For example, the original theory proposed by
\cite{lyndenbell} to predict the properties of 
virialized states made both the assumption that these states 
have the same energy and mass as the initial conditions, and
that the dynamics is purely mean field (i.e. governed by the 
Vlasov-Poisson equations). Although this theory is believed to
be inadequate\footnote{For recent studies see e.g. 
\cite{arad, arad+lyndenbell_2005}, and also that of 
\cite{levin_etal_2008}.}
--- and in any case it may not be applicable to
the particular set of cold initial conditions we study ---
such assumptions are usually made in theoretical approaches
to understanding these states. It is thus important to understand
the extent to which they are valid. Interestingly we find 
in our simulations that the {\it shape} of the density profiles 
of the relaxed systems appear to be very robust, i.e., 
independent of the initial conditions, despite the fact that 
their global macroscopic parameters, notably their mass and energy, 
vary with $N$.

In this article we address these questions primarily using 
the very restricted set of initial conditions given by distributing
$N$ particles randomly in a spherical volume, and ascribing
zero velocity, i.e., a sphere of cold matter with Poissonian density
fluctuations. The simplicity of these initial conditions is
that they are characterized by the single parameter $N$.
On the other hand, they present an intrinsic numerical 
complexity:  while the problem is well defined for any
finite $N$, it tends formally, as $N \rightarrow \infty$, to the
exactly uniform spherical collapse which leads to a density
singularity at a finite time.  This makes it expensive to integrate
numerically for increasing $N$. Indeed for this reason many authors have
excluded it from numerical studies, focusing instead on spherical
models with non-trivial inhomogeneous distributions (e.g. radially
dependent density) and/or significant non-zero
velocities \citep{vanalbada_1982, mcglynn_1984, villumsen_1984,
aguilar+meritt_1990, theis+spurzem_1999, merrall+henriksen_2003,
roy+perez_2004, boily+athanassoula_2006}. As we will see, it 
is precisely the fact that the
system collapses by a large factor, allowing particles to reach
very large velocities, that leads to the very significant energy 
ejection which we focus on. We will discuss, in the part
of the paper on the mean-field approximation, some other very
specific initial conditions. We will also consider in our conclusions 
the extent to which our study may be pertinent to other
initial conditions. 

We note also
that these initial conditions are the most evident discretisation
of the exactly uniform spherical collapse model, which is a 
reference point in the theory of non-linear evolution of
self-gravitating systems in astrophysics and cosmology (see, e.g.
\cite{halo}). The 
assumption usually made in using this model to predict the 
behaviour of physically relevant systems (e.g. in cosmology, 
via the formalism of Press and Schecter) is precisely that
the total initial mass and energy are virialized. Our study
shows that this is not a good approximation for this class
of ``almost'' uniform spherical collapses. 

Despite the numerical difficulties associated with these
initial conditions, several extensive studies of them have,
however, been reported  in the literature, most notably 
a detailed study by \cite{aarseth_etal_1988}, and a more
recent one reported in \cite{boily_etal_2002}. 
We will discuss these works in greater
detail below, and use some of their results. Both works focus 
on how the singular collapse of the infinite $N$ limit is regulated
when there are a finite number of particles, and in particular the
scaling observed in numerical simulations for the minimal size
of the system\footnote{In the context motivating the study 
of  \cite{aarseth_etal_1988} the ``points'' are actually masses 
with extension (e.g. proto-stars) and the central question 
the authors wish to address is whether these masses survive
or not the collapse of a cloud of which they are the constituents.}.  
\cite{boily_etal_2002} also considers a wider class of non-spherical
collapses, finding results to which we will return briefly 
in our conclusions.  Our study can be seen as an extension of, and is
complementary to, these studies, with which our results are in
agreement. In more recent work 
\cite{morikawa_nongauss, morikawa_virial} have also studied
collapse from  such initial conditions, focussing on the 
properties of the virialized state. Although 
they do not consider the ejected energy 
quantitatively, these authors do remark on its potential importance
in open systems. Some of the works cited above on collisionless
relaxation (notably \cite{vanalbada_1982, aguilar+meritt_1990, roy+perez_2004})
do include some quite cold uniform spherical initial conditions,
with initial virial ratios of order $0.1$, and note that there 
is significant mass ejection for this case.  
For the relatively small number of particles considered in these
cases the energy ejection is, however, modest and does not strongly
modify the final state, and the dependence of this quantity on
particle number has not been explored\footnote{
\cite{aguilar+meritt_1990} actually discount a 
possible systematic dependence on $N$
in a footnote on page 36.}.
We note also the studies in \cite{david+theuns_1989,
theuns+david_1990}, which explicitly discuss and model 
the ejection of particles (``escapers'') from pulsating 
spherical systems. 
 
We will describe here not only the dependence of the energies,
and various other quantities, on $N$, but also detail the physical
mechanism which leads to the ejection of mass. The probability of 
ejection turns out to be closely correlated with particles' initial 
radial positions,  with essentially particles initially in the 
outer shells being  ejected. The reason for this correlation is 
simply that these particles which are initially near the outer 
boundary systematically lag (in space and time) with respect
to their uniform spherical collapse trajectories more than 
those closer to the centre\footnote{We note that this formation
of a ``core-halo'' structure during collapse is remarked on and
briefly discussed by \cite{aarseth_etal_1988}. The link
to mass/energy ejection are not, however, discussed in this
article, or in \cite{boily_etal_2002}.}. These particles then 
gain the energy leading to their ejection in a very short 
time around the collapse 
time, as they pass through the  time-dependent potential of the 
particles initially closer to the centre, which have already collapsed 
and ``turned around''. This lag, which has a very non-trivial
dependence on the initial radial position, is manifestly a physical
effect coming from the boundary: fluctuations to uniformity evolve
differently depending on their position with respect to the boundary.
Indeed the lag which leads to the mass ejection first develops
at the outside of the system and propagates into the volume during
the collapse phase. The importance of the finite size of the system
in this respect is in contrast to the physics required to understand
the scalings observed by \cite{aarseth_etal_1988} and
\cite{boily_etal_2002}. These can be
understood by analyzing only the growth of fluctuations in the 
collapsing system in the approximation that it is of arbitrarily
large size, which corresponds to the case of a matter dominated 
contracting universe. Nevertheless, as we will explain,  we can use these 
arguments also to understand the scaling we observe of the 
ejected energy once the lag is assumed to be produced at collapse
by the mechanism desribed above. While this is highly consistent
with a description of the ejection phase which is clearly 
mean-field like --- the lagging particles propapating through 
the time dependent potential of the rest of the mass --- it is 
not, as we discuss, evident whether this is true of the physics 
involved in the development of the lag (and therefore determination 
of the ejected mass). In the final section we address more precisely
what this mean field limit is, and how one should extrapolate 
numerically towards it. We then report some numerical tests on
appropriate modifications of the initial conditions, which probe
such convergence. While we establish that there is indeed good
evidence for convergence, there are still measurable fluctuations
due to non-mean field effects in macroscopic quantities such 
as the ejected energy at the largest particle numbers we
have simulated. 

The article is organized as follows. In the next section we recall
the predictions for the $N$ dependences of various quantities
which follow from the uniform and perturbed spherical collapse
model. In the following section we describe our simulations and
principle results concerning the $N$ dependence of the final
state produced by violent relaxation. In 
Sect.~\ref{Mechanism of mass and energy ejection} we investigate
in detail the mechanism which leads to the ejection of mass 
and energy which we have observed. In Sect.~\ref{mean-field}
we discuss the validity of the Vlasov-Poisson limit in
describing the evolution we have studied. Finally we
give a summary of our findings and conclusions
in the last section.

\section{Scalings in the perturbed spherical collapse model} 

In this section we recall basic results on the exactly
uniform spherical collapse model, and then the scalings of
physically relevant quantities (minimum attained radius,
time of collapse, velocities at collapse etc.) which are
obtained from a straightforward analysis of the evolution
of perturbations to this model. 

\subsection{Uniform spherical collapse} 

The radial position $r(t)$ of a test particle in an (idealized) exactly 
uniform spherical distribution of purely self-gravitating matter of initial 
density $\rho_0$ and initially at rest (at time $t=0$) is simply
given by the homologous rescaling 
\be
r(t) = R(t) r(0)
\ee
where the {\it scale factor} $R(t)$ may be written in 
the standard parametric form
\bea
&& 
R(\xi) = \frac{1}{2} (1 + \cos(\xi))
\\ \nonumber 
&&
t(\xi) 
= \frac{\tau_{scm}}{\pi} \left( \xi + \sin(\xi) \right)
\;, 
\label{scm2}
\eea and
\be 
\label{tauscm}
\tau_{scm} \equiv \sqrt{\frac{3\pi}{32 G \rho_0}} \;.
\ee 
At the time $\tau_{scm}$ 
the system collapses to a singularity. It will be useful in what
follows to recall how the physical quantities diverge 
close to this singularity. Taking $\xi = \pi -\epsilon$ 
and expanding to leading order in $\epsilon$ gives
$(t-\tau_{scm}) \sim \epsilon^3$, from
which it follows that
\be
\label{asymp-R}
R(t) \sim [t-\tau_{scm}]^{2/3}
\ee
and therefore the test particle velocities $v(t)$,
proportional also to the initial radius $r(0)$, 
scale as
\be
\label{asymp-v}
v(t) \sim [t-\tau_{scm}]^{-1/3}.
\ee

\subsection{Perturbed spherical collapse} 

In this exactly uniform limit the evolution
is independent of the size of the system, 
and the ``particles'' do not see its finite size 
(until the collapse).  These solutions are thus
formally valid as the radius of the system tends
to infinity, and indeed they are precisely those 
for the evolution of an infinite contracting universe 
containing only matter, which are known to 
coincide with those derived in general relativity. 
The last equations given above then correspond to 
the behaviours in a contracting Einstein de Sitter 
(EdS) universe obtained when the curvature (corresponding 
to the initial cold start) is neglected. 

The system we study here --- $N$ randomly
placed particles in a spherical volume --- can be
treated, up to some time 
and at sufficiently large scales,
as a perturbed version of this uniform limit. While in general 
the associated perturbations about uniformity will be 
expected to evolve in a way which will be sensitive to 
the finite size of the system, an approximation one can 
make is that such effects are negligible, i.e. to treat 
the perturbations as if they evolve also in an infinite 
contracting system. Quite simply this means we neglect
the effect of the boundaries on the evolution of the
density perturbations.  

In the manner standard in
cosmology (for the case of an expanding universe)
one can then consider the fluid limit for the system
and solve the appropriate equations perturbatively
(see e.g. \cite{peebles}). In the eulerian formalism 
this gives, at linear 
order, a simple equation for $\delta (\bx)$, the 
density fluctuation (with respect to the mean density):
\be
\label{deltaevol1}
\ddot \delta + 2 H \dot \delta - 4 \pi  G \rho_0 \delta =0
\ee
where $H(t)=\dot{R}/R$ (dots denotes derivatives with respect to
time) is the contraction (``Hubble'') rate. These equations
are derived in ``comoving'' coordinates $\bx=\br/R(t)$, where
$\br$ are the physical vector positions. Note that 
\be
R \,\dot{\ve x}=\dot{\ve r} - \dot R\, \ve x \,,
\ee
i.e., $R(t) \dot{\bx} (t)$ is the velocity of 
the particle minus the velocity it would have in the limit 
of uniformity (in the cosmological context, the ``peculiar'' 
velocity with respect to the ``Hubble flow'').

It is straightforward (see Appendix A) to solve Eq.~(\ref{deltaevol1})
by rewriting is as an equation for $\delta (R)$. 
In the limit $R \ll 1$, 
\be
\label{limita0}
\delta(R) \sim R^{-3/2}\,
\ee
i.e., 
\be
\label{delta-limita0}
\delta (t) \sim [t-\tau_{scm}]^{-1} \,.
\ee
This is simply the usual decaying mode of the expanding 
EdS universe,  which becomes the dominating growing mode in the 
contracting case.

A much more detailed 
analytic treatment of perturbations to the spherical 
collapse model in the fluid limit, for the finite
system, may be found in \cite{aarseth_etal_1988}. 
We give only the above results here, for the infinite radius limit,
as they are the only ones we will make use of below.

\subsection{Predictions for scalings}
\label{Predictions for scalings}

The singular behaviour of the spherical collapse is regulated by
the fluctuations present at any finite $N$ in the initial conditions
we study. A simple estimate of the scale factor $R_{min}$ at which 
one expects the spherical collapse model to break down completely
may be obtained by assuming that this will occur when fluctuations 
at some scale (e.g. of order the size of the system) go non-linear. 
For Poisson distributed particles we have a mass variance (see e.g. 
\cite{book})
\be
\sigma^2(r) = \frac{\left( \langle N(r) - \langle N(r) \rangle \right)^2}{\langle N(r) \rangle^2}
\propto \frac{1}{\langle N(r) \rangle} \;,
\label{poisson-variance}
\ee
where the angle brackets denote an ensemble average over realizations.
Using this as the initial normalisation of the density fluctuations,
and the growth given in Eq.~(\ref{limita0}), we can infer
\be
\label{collapse-scaling}
R_{min} \propto N^{-1/3} \;.
\ee
Using the scalings in Eqs.~(\ref{asymp-R}) and (\ref{asymp-v})
it then follows that the maximum time $t_{max}$ until which
the spherical collapse model is a reasonable approximation is
expected to scale as 
\be
\label{tmax-scaling}
[t_{max} - \tau_{scm}] \propto N^{-1/2} \;,
\ee
and the maximal infall velocity at any given radius 
as
\be
\label{vmax-scaling}
v_{max} \propto N^{1/6} \;.
\ee

Note that we do not need to make explicit any scale in the
argument to obtain this result: the scaling with $N$ is obtained
simply by taking the criterion that {\it some} amplitude
is reached by the fluctuations at {\it some} fraction of the system
size. In fact the result can be obtained on purely dimensional
grounds: given that we are neglecting the finite size of the
system, the only length scale in the problem --- as gravity
itself furnishes no scale --- is given by 
the mean interparticle distance $\ell \propto N^{-1/3}$.
Clearly then the length scale determined when $R_{min}$
is combined with the size of the system
must scale in this way, if the physical processes determining
it are indeed independent of the size of the system.

Both \cite{aarseth_etal_1988} and \cite{boily_etal_2002} use 
simple, but equivalent, variants of the above argument to obtain these
same scalings. In particular they can be formulated in
terms of the spread in the collapse time of approximately
spherical overdense and underdense regions, with initial
amplitude fixed by Eq.~(\ref{poisson-variance}). Or,
alternatively, one can analyse the scaling of the 
pressure associated with growth of the peculiar velocities
which must compensate the inward gravitational pressure 
to stop the collapse. One of the central findings of
both \cite{aarseth_etal_1988} and \cite{boily_etal_2002} 
is that the scaling
Eq.~(\ref{collapse-scaling}) is in fact observed, and 
we will verify again in our simulations that this is 
indeed the case. 

The fact that these scalings are observed --- and thus that
the underlying approximation of (i) neglecting the boundaries
of the system, and (ii) treating the fluctuations in the 
fluid limit, appears to work well --- makes it instructive
to compare collapsing spheres with different numbers of
initially Poisson distributed particles at appropriately 
rescaled times. Let us consider two such spheres, with
say $N_1$ and $N_2 > N_1$ particles, as if they were
just two spheres of different radii ($R_1$ and 
$R_2=R_1(N_2/N_1)^{1/3}$) evolving in an infinite 
collapsing universe with initially Poisson distributed 
points. In the approximation that the perturbations 
grow as given by linear fluid theory [i.e. as in
Eq.~(\ref{limita0}) above], the sphere with $N_2$
points should be exactly equivalent, up to
transients due to the cold start, to that with
$N_1$ points at the initial time after the evolution
of the scale factor brings its initially smaller
density fluctuation (of amplitude $\propto N_2^{-1/2}$)
to the initial larger amplitude ($\propto N_1^{-1/2}$)
of the sphere with $N_1$ particles. This gives precisely
the rescaling inferred above, i.e., the different spheres
should more generally --- and not just at the time 
corresponding to $R_{min}$ --- be equivalent at scale factors 
rescaled in proportion to $N^{1/3}$. Deviations from equivalence
in these rescaled time variables, on the other hand, must
arise from physical effects not described by this approximation
of neglecting the finite size of the system in its evolution
up to collapse. We will thus compare below the different $N$ 
systems at these different scales factors, and discuss 
to what extent the observed scalings of the ejected 
mass and energy can be understood in this simple approximation.

\section{Cold Spherical collapse: Basic results for Poissonian fluctuations}
\label{Basic results}

We describe in this section numerical results for evolution from initial
conditions in which $N$ particles are distributed randomly in a sphere,
and given zero initial velocity. Besides reproducing known results for 
various quantities --- notably the minimal collapse radius (or maximal 
potential energy), and the final profiles of the virialized structures --- we
present also our results for the $N$ dependence of the ejected mass
and energy. 

\subsection{Initial conditions and choice of units}

Table~\ref{table-one}  shows the names of the different simulations and the 
associated particle number $N$. The $i$-th realisation
of an $N$ particle configuration is thus labelled $PN-i$.
The parameter $\ell$ is the {\it mean interparticle separation}
in the initial configuration, defined as 
$\ell \equiv (3V/4\pi N)^{1/3}=R/N^{1/3}$
where $V$ is the volume of the sphere of radius $R$.
The {\it unit of length}, here and throughout the paper, is the
{\it diameter of the initial sphere}, i.e., $R=0.5$. 
The parameter $m$ is the (identical) mass of the particles. 
The {\it unit of mass}, here and throughout the paper, is 
chosen {\it so that the initial mass
density $\rho_0$ is unity}, i.e., $\rho_0 = mN/V=6mN/\pi=1$.
Finally we take our {\it unit of time equal to} $\tau_{scm}$,
the uniform spherical collapse time as defined in Eq.~(\ref{tauscm}) 
\footnote{In these units therefore
$G=3\pi/32$. In physical units $\tau_{scm}$ is approximately
$2100$ seconds if $\rho_0$ is $1$g/cm$^2$.}.

The parameter $\varepsilon$ is the softening parameter introduced
in the numerical integration, which we will discuss at length below.

\begin{table}
\begin{tabular}{ccccc}
\hline
Name     &   N    &       $\ell$    &      $\varepsilon/\ell$ &       $m$       \\
\hline
P512-1     &   512  &       0.63E-01  &      0.05  &       0.10E-02   \\
P512-2   &   512  &       0.63E-01  &       0.05 &       0.10E-02   \\
P512-3   &   512  &       0.63E-01  &       0.05  &       0.10E-02   \\
P1024-1    &  1024  &       0.49E-01  &       0.065  &       0.51E-03   \\
P1024-2  &  1024  &       0.49E-01  &       0.065 &       0.51E-03   \\
P1024-3  &  1024  &       0.49E-01  &       0.065  &       0.51E-03   \\
P2048-1    &  2048  &       0.39E-01  &       0.082  &       0.26E-03   \\
P2048-2  &  2048  &       0.39E-01  &       0.082  &       0.26E-03   \\
P2048-3  &  2048  &       0.39E-01  &       0.082  &       0.26E-03   \\
P2048-4  &  2048  &       0.39E-01  &       0.082  &       0.26E-03   \\
P2048-5  &  2048  &       0.39E-01  &       0.082  &       0.26E-03   \\
P4096-1    &  4096  &       0.25E-01  &       0.10  &       0.13E-03   \\
P4096-2  &  4096  &       0.25E-01  &       0.10  &       0.13E-03   \\
P4096-3  &  4096  &       0.25E-01  &       0.10  &       0.13E-03   \\
P8192-1    &  8192  &       0.25E-01  &       0.13  &       0.64E-04   \\
P8192-2  &  8192  &       0.25E-01  &       0.13  &       0.64E-04   \\
P8192-3  &  8192  &       0.25E-01  &       0.13  &       0.64E-04   \\
P16384   & 16384  &       0.20E-01  &       0.16  &       0.32E-04   \\
P32768-1   & 32768  &       0.16E-01  &       0.2  &       0.16E-04   \\
P32768-2  & 32768  &       0.16E-01  &       0.2  &       0.16E-04   \\
P32768-3  & 32768  &       0.16E-01  &       0.2  &       0.16E-04   \\
P32768-4  & 32768  &       0.16E-01  &       0.2   &       0.16E-04   \\
P32768-5  & 32768  &       0.16E-01  &       0.2  &       0.16E-04   \\
P65536   & 65536  &       0.12E-01  &       0.25  &       0.80E-04   \\
P131072  &131072  &       0.99E-02  &       0.33  &       0.40E-05   \\
P262144  &262144  &       0.78E-02  &       0.41  &       0.20E-05   \\
\hline
\end{tabular}
\caption{Details of the simulations: $N$ is the number of randomly
distributed particles in a sphere of diameter taken equal to unity,
$\ell$ is the average interparticle separation (see text for
definition),
$\varepsilon$ is the softening length in the gravitational
potential and $m$ is the particle mass, in units in which the
mean mass density is unity. 
\label{table-one} }
\end{table}

\subsection{Numerical code and parameters}
 
We have performed numerical simulations using the publicly
available code GADGET2 \citep{gadget, springel_2005}. This code, which is
based on a tree algorithm for the calculation of the gravitational
force, allows one, in particular, as desired here, to perform 
simulations of a finite system with open boundary conditions.  The 
two-body potential used 
is exactly the Newtonian potential for separations greater than
the 
softening length $\varepsilon$, and modified at smaller scales
to give a force which is attractive everywhere and vanishing
at zero separation\footnote{The exact expression for the 
smoothing function may be found in \cite{gadget_paper}.}.

The values of the ratio $\varepsilon/\ell$ given in 
Table~\ref{table-one} correspond in fact to the single fixed
value $\varepsilon=0.0028$, i.e. the smoothing in this set of
simulations is fixed in units of the initial size of the 
the system. Given that our aim here is to reproduce as closely
as possible the evolution of the particle system 
(i.e. without smoothing), we will test carefully 
for the dependence of our results on the choice of 
$\varepsilon$, and, specifically, for their stability
when $\varepsilon$ is decreased compared to the value
chosen in this set of simulations. As this discussion
is closely related to the issue of the validity of the
mean field (Vlasov Poisson) approximation to the dynamics
of the system, we will present these results in 
Sect.~\ref{mean-field}, where we discuss this question.
For now we note simply that if such a mean field approximation
is valid, one expects that it should be sufficient that
$\varepsilon$ be significantly smaller at all times than
the length scales relevant to this dynamics. We will
see that the collapse phase leads to a minimal length
scale of the structure of order the scale $\ell$ (and, as established by 
\cite{aarseth_etal_1988} and \cite{boily_etal_2002}, proportional to this
scale). Given the values
for $\varepsilon/\ell$ shown in Table~\ref{table-one},
it follows that the chosen $\varepsilon$ is, in
the largest $N$ simulation, still smaller than this 
minimal collapse scale.

The large compression which the system undergoes as it collapses
requires that particular care be taken in choosing the numerical
parameters which control the precision on the force calculation
and time stepping. Given the physics of the system, we have
used prescriptions for parameter choices in which we divide 
the entire range of
times in three different phases. During the first ($0 \le t \le 0.95
\tau_{scm}$) and the third ($1.14 \tau_{scm}\le t \le 3.5 \tau_{scm}
$) we have used a relatively large time step (of the order of $5
\times 10^{-4} \tau_{scm}$) while in the collapse phase ($0.95
\tau_{scm} \le t \le 1.14 \tau_{scm}$ sec) we have used a more
accurate time step (of the order of $5 \times 10^{-5} \tau_{scm}$).
We have performed several tests to check the 
stability of the results given below, at the relevant level of 
numerical precision. We have in all cases quantified carefully 
the conservation of the total energy, which experience has 
shown to be a very sensitive quantity for monitoring the 
accuracy of the simulation \citep{aarseth_book}. Shown, for example,
in Fig.~\ref{fig_etot} is the temporal evolution of the total
energy $E_T$ normalised to the initial energy $E_0$ in
three of the simulations in Table~\ref{table-one}. 
\bef 
\centerline{\includegraphics*[width=8cm]{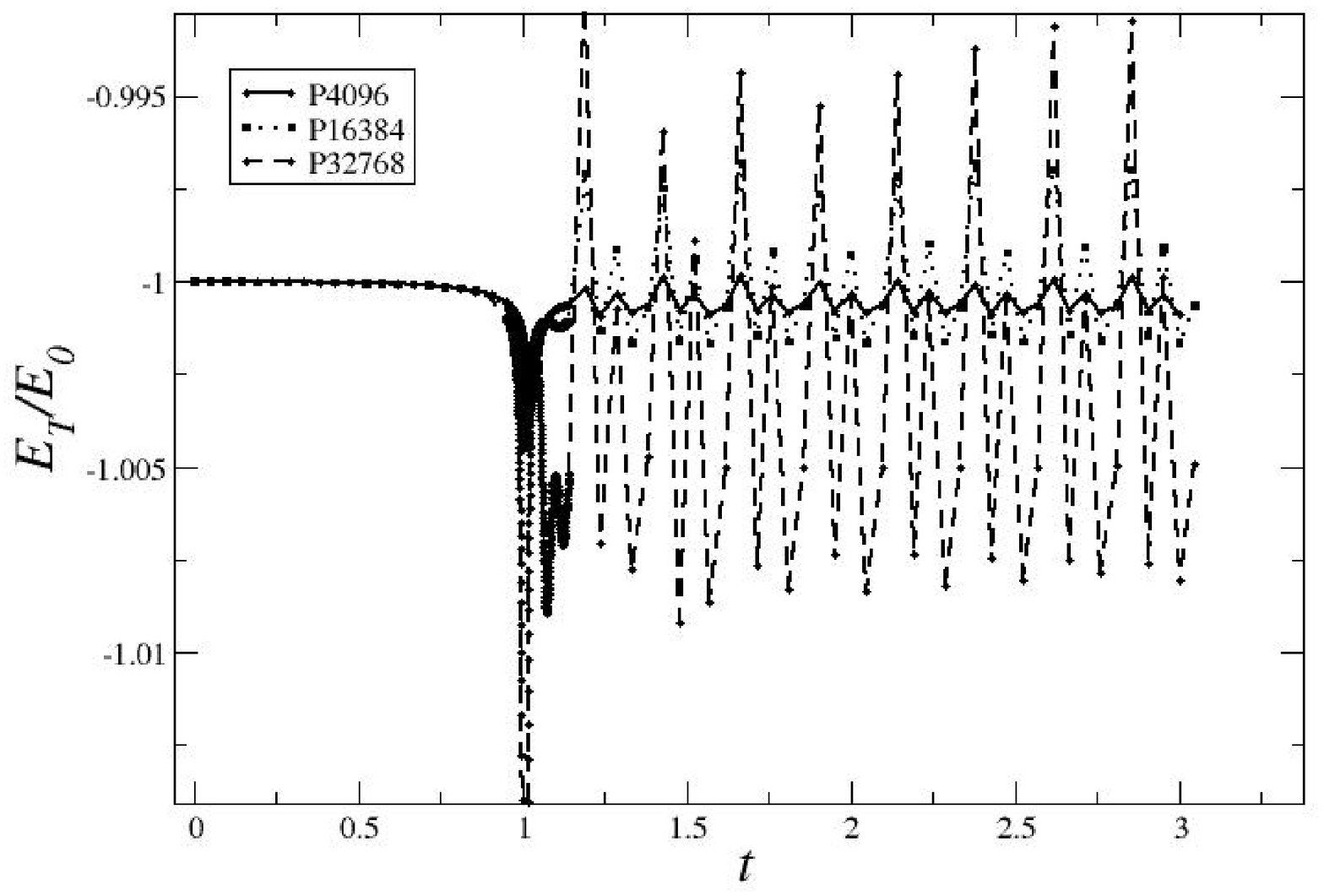}}
\caption{Total energy in the indicated simulations as a function of time. 
\label{fig_etot} }
\eef
For larger $N$ the energy is conserved less well, but 
the fluctuations are of less than one percent
for all but the largest $N$. For the very largest 
$N$ simulation in Table~\ref{table-one} the fractional error
in the total energy reaches about $5 \%$ during the collapse
phase. However, in this case and all the simulations, these 
variations in the energy 
are fluctuations about the total energy which do not lead to 
any measurable drift in the average value, which would be
indicative of systematic error which propagates in time
\footnote{The leap-frog time-integration method used by
GADGET2 has, due to its
simplectic nature, the property that it conserves extremely 
well --- despite local fluctuations in time ---  the 
total energy of the system. For a detailed discussion see 
\cite{springel_2005}.}. 

A further test of our simulations is their consistency with
previous results reported in the literature for the same system, 
which we will discuss in the next subsection. We will then 
present our results for the dependence on initial conditions
(i.e. on $N$) of the final virialized state.

\subsection{Basic qualitative and quantitative results}

Qualitatively the behaviour of this specific kind of cold
collapse initial conditions is well known, and has been studied
in detail in \cite{aarseth_etal_1988} and \cite{boily_etal_2002}: the 
system collapses
following approximately the spherical collapse solution until
it reaches some minimal size, at a time of order $\tau_{scm}$.
It then ``turns around'' and rapidly (i.e. in a time 
considerably shorter than $\tau_{scm}$) settles down to a 
quasi-stationary virialized state. This is illustrated 
in Fig.~\ref{fig_virial} which shows, for the simulations 
indicated (cf. Table~\ref{table-one}), the evolution of the ratio
\be
\label{virial}
b(t) = \frac{2K^n}{W^n}
\ee
where $K^n$ is the kinetic energy of the particles with
negative energy, and $W^n$ the potential energy associated
with the same particles. The ratio represents the virial
ratio of the entire system at early times, and that of 
the bound particles at late times.

\bef 
\centerline{\includegraphics*[width=8cm]{Fig2.eps}}
\caption{The virial ratio for the particles with negative energy 
[see Eq.(\ref{virial})] as a function of time for different
simulations. In the insert panel is shown a zoom of the behaviour 
after the collapse. 
\label{fig_virial}
}
\eef
A typical evolution of the radial density profile is shown in 
Fig.~\ref{fig_dpt}: until close to the maximal collapsed 
configuration, it maintains, approximately, the top-hat 
form of the original configuration and then in a very
short time changes and stabilizes to its asymptotic form.
We will discuss below in further detail this latter form 
and will also study in detail the deviations from 
the original form during the collapse phase. 
\bef
\centerline{\includegraphics*[width=8cm]{Fig3.eps}}
\caption{Density profile of particles with 
negative 
total energy at different times during the collapse, for the P65536 simulation.
\label{fig_dpt}
}
\eef

Studies of the $N$ dependence of the evolution have focussed mainly
on the question of how the minimal size reached by the structure 
depends on this parameter.
This minimal radius $R_{min}$ may be defined in different ways,
e.g., as the minimal value reached by the radius, measured from
the centre of mass, enclosing $90\%$ of the mass. Alternatively
it can be estimated as the radius inferred from the potential
energy of the particles, the minimal radius corresponding to
the maximal negative potential energy\footnote{Spherical
symmetry is maintained up to a very good approximation until
this time.}. The behaviour of $R_{min}$, 
determined by the first method, as a function of $N$ is 
shown in Fig.~\ref{fig_rmin}. This observed behaviour is in 
good agreement with the prediction of the scaling 
Eq.~(\ref{collapse-scaling}), in accord with the findings
of \cite{aarseth_etal_1988} and \cite{boily_etal_2002}\footnote{The latter
paper finds that this result extend to $N$ of order $10^7$,
an order of magnitude beyond the largest $N$ we report here.}. 
We note that the factor of proportionality between $R_{min}$
and $\ell$ is in fact almost unity (as in our units
$\ell=0.5/N^{1/3}$).
\bef
\centerline{\includegraphics*[width=8cm]{Fig4.eps}}
\caption{ Behaviour of the minimal radius $R_{min}$ attained,
determined as described in text,  as a function of $N$.
The solid line is the best fit to the prediction 
of Eq.~(\ref{collapse-scaling}).
\label{fig_rmin}
}
\eef

In Fig.~\ref{fig_wmin}
the absolute value $|W_{min}|$ of the potential energy at its minimum
value, as a function of $N$. We again observe an excellent fit 
to the predicted behaviour.
\bef
\centerline{\includegraphics*[width=8cm]{Fig5.eps}}
\caption{The absolute value of the minimal value reached by the 
potential energy (i.e. at the time $t=t_{max}$), as a function of $N$.
\label{fig_wmin}
}
\eef

\subsection{$N$ dependence of ejected mass}

Our focus in this paper is on the presence of an ejected component 
of the mass, and energy, which has not been closely examined in
these previous works. One of the features of the collapse and
virialization is indeed that while all particles start with a 
negative energy, a finite fraction end up with a positive 
energy. Given that they move, from very shortly after the collapse,
in the essentially time independent potential of the 
virialized (negative energy) particles, they escape from 
the system. Indeed at the times at which we end our 
simulations all such particles are very far from the 
collapsed region and move outward with almost constant energy,
with a negligible probability of an encounter which 
could stop their escape to arbitrarily large distances.

This transfer of energy, which we will examine in detail
below, occurs in a very short time around the maximal
collapse, and depends on $N$. Both these facts
can be seen in Fig.~\ref{fig_fpt}, which shows a plot
of the fraction $f^p$ of the particles with positive
energy at any time. 
\bef
\centerline{\includegraphics*[width=8cm]{Fig6.eps}}
\caption{ The behaviour of $f^p(t)$, the fraction of the particles
with positive energy, as a function of time for two
  different simulations. A dependence on
  the number of particles is manifest.
\label{fig_fpt}
}
\eef

Note that, while our simulations indicate that $f^p$ attains
a well defined asymptotic value, this is not expected to be
the truly asymptotic value of this quantity, or of any of
the other quantities we consider below: on much longer
time scales than those considered here
(i.e. the times characteristic of the violent relaxation
of the system) further particles may gain energy and 
escape from the system, notably by two body 
encounters (see e.g. \cite{binney}). If the system
is ergodic, simple considerations based on the 
microcanonical entropy (see e.g. \cite{Padmanabhan:1989gm}) 
imply that at asymptotically long times, and for an 
unsmoothed gravitational potential, the particles will
tend to a configuration in which there is a single pair 
of particles with arbitrarily small separation, and the 
rest of the mass is in an ever hotter gas of free particles
(see e.g. \cite{aarseth_1974}). 
When the potential is smoothened, as in our simulations
here, we would expect on such very long time scales to
obtain a final state with some part of the mass bound
and which depends strongly on this smoothing. We will,
however,  not explore this temporal regime here.

The $N$ dependence of the asymptotic value of $f^p$ we identify 
in this way is shown in Fig.~\ref{fig_fpos}. We note that, although 
determinations in different realizations of a given system with 
the same number of particles fluctuate, there appears to be
a very slow, but systematic, increase as a function of $N$,
which is reasonably well fit by
\be
\label{fposlog}
f^p(N) \approx  a+ b \log(N)\;,
\ee
where $a=0.048$ and $b=0.022$. Alternatively it can be fit 
quite well (in the same range) by a power law 
$f^p \approx 0.1 N^{0.1}$.

\bef
\centerline{\includegraphics*[width=8cm]{Fig7.eps}}
\caption{Behaviour of the fraction of ejected particles as a function
of the total number of particles in the system. The solid line
is the phenomenological fit given by Eq.~(\ref{fposlog}).
\label{fig_fpos}
}
\eef

\subsection{$N$ dependence of ejected energy}

While the factor of the mass ejected varies, at most, very slowly
with $N$, the energy it carries away has a much stronger dependence
on $N$. In Fig.~\ref{fig_kinetic_energy} is shown, for example, the 
behaviour of the total kinetic energy as a function of time 
in two simulations with different $N$. Not only the time at which
the maximum is reached, but also the final value of the energy, 
change clearly\footnote{The common evolution of both curves is
simply that predicted by the spherical collapse model,  which
leads to a divergence of the kinetic energy at $\tau_{scm}$.}. 
\bef
\centerline{\includegraphics*[width=8cm]{Fig8.eps}}
\caption{Behaviour of the total kinetic energy for 
two simulations with different number of particles. 
\label{fig_kinetic_energy}
}
\eef
We consider now carefully this dependence, and the related one of
the potential energy of the different components.

In general the total energy, which is equal to the initial energy
$E_0$, may be written as
\be E_0 = W + K = W^p + W^n + K^p + K^n + W^{p/n} 
\ee 
where $W$  ($W^p$, $W^n$) and $K$ ($K^p$, $K^n$) 
are the total potential and kinetic energy of the 
particles  (with positive energy, with negative energy)
at the time $t$,  and $W^{p/n}$ is the potential energy 
associated with the interaction of the particles with 
positive energy with those of negative total energy. 
Note that $E_0$ is simply, up to a correction from 
fluctuations due to the particle discreteness, the 
gravitational potential energy of a uniform ball of 
radius $r_0$, i.e., 
$E_0= W(t=0)= - \frac{3}{5} \frac{GM^2}{r_0}$, where 
$M=mN$ is the mass of the ball (see e.g. \cite{binney}). 

Shown in 
Fig.~\ref{fig_energyP65536} are the evolution of these quantities\footnote{The energies here and in
subsequent figures, unless specified, are given in units 
of the absolute value of the initial total energy $E_0$.}. 
The maximum of the kinetic energy $K$ 
corresponds evidently to the minimum of the potential energy $W$.
We see also that $W\approx W^n$, with only a very slight deviation
around the time of collapse. This corresponds simply to the
expulsion of the positive energy particles, which makes
both $W^p$ and $W^{p/n}$ completely negligible once
these particles are far from the remaining bound part
of the system.
\bef
\centerline{\includegraphics*[width=8cm]{Fig9.eps}}
\caption{{\it Upper panel:} behaviour of the kinetic and potential
energy, normalized in units of the absolute value of the initial 
energy, for all
  particles and for particles with negative total energy
  as a function of time for the P65536 simulation.
{\it Lower panel:} The same for particles with positive total energy.
\label{fig_energyP65536}
}
\eef

Once the collapse phase is over, we therefore have to an
excellent approximation
 \be
\label{energyapprox}
E_0 = W^n + K^p + K^n  \;.
\ee
Further, since the bound particles rapidly virialize into a 
quasi-stationary state, we have
\be
\label{energyapprox2}
2 K^n + W^n =0 \;.
\ee

The three non-zero energies  $K^p$, $K^n$ and $W^n$ can therefore
be expressed in terms of the initial (potential) energy $E_0$
and a single a priori unknown quantity, e.g., the ejected energy
$K^p$.

Note that it follows from Eq.~(\ref{energyapprox2}) and 
Eq.~(\ref{energyapprox}) that
\be
\label{energyapprox3}
E_0 = K^p + \frac{W^n}{2} \;,  
\ee 
so that for the case we are studying, in which $E_0 <0$,  these 
considerations do not on their own imply the necessity for 
energy ejection, i.e., one can satisfy
Eq.~(\ref{energyapprox3}) with $K^p=0$. For $E_0>0$, however,
the formation of a bound virialized stationary state requires 
such ejection.

In Fig.~\ref{fig_kpos} 
we show the observed behaviour in our simulations of the total 
kinetic energy $K^p$ of the ejected particles as a function 
of $N$. Two fits are shown: one is to a single power-law
behaviour for $K^p$ itself, the other to the behaviour
which would be observed, using  Eq.~(\ref{energyapprox3}), 
if $W^n$ scaled as $W^{min}$ observed above. Thus in 
these curves we see a behaviour which is close to, 
but clearly different from, the simple scaling
observed for $W^{min}$.
\bef
\centerline{\includegraphics*[width=8cm]{Fig10.eps}}
\caption{ Behaviour of the total kinetic energy of the ejected particles
(normalized to initial total energy)  as a function of the  number of
particles in the system.
The dashed line is the best fit to a single power-law
behaviour for $K^p$ itself, the solid line a fit
to Eq.~(\ref{energyapprox3}) with  $W^n \propto N^{1/3}$,
i.e. with $W^{n}$ scaling exactly as $W^{min}$ observed above.
\label{fig_kpos}
} 
\eef 

Shown in Fig.~\ref{kposfpos} is instead the ratio $K^p/f^p$, i.e.,
the kinetic energy {\it per unit ejected mass}, as a function of $N$.
\bef
\centerline{\includegraphics*[width=8cm]{Fig11.eps}}
\caption{Observed behaviour of the ratio $K^p/f^p$ for the
set of simulations in Table~\ref{table-one}. 
\label{kposfpos}
}
\eef
We observe that the simple behaviour $K^p/f^p \propto N^{1/3}$
provides a very good fit, with less scatter than in the
previous plot. After we have discussed below in detail the
mechanism of mass/energy ejection, we will give a simple
scaling argument which accounts for this behaviour.

\subsection{$N$ dependence of density profiles}

The radial density profile of the virialized structure
formed by the bound particles after the collapse 
may be well fit in all our simulations by the 
functional form
\be
\label{dpscm1}
n(r) = \frac{n_0}{\left(1+\left(\frac{r}{r_0}\right)^4\right)} \;.
\ee
While the previous studies of exactly cold uniform initial 
conditions do not report results for this quantity, we observe
that it is in agreement with that found for cold (i.e. low
initial virial ratio) initial conditions in several previous studies 
(see e.g. \cite{henon_1964, vanalbada_1982, roy+perez_2004}).

\bef
\centerline{\includegraphics*[width=8cm]{Fig12.eps}}
\caption{Behaviour of the parameter $r_0$ as a function of the number of particles
in the simulation. 
\label{fig_r0N}
}
\eef
\bef
\centerline{\includegraphics*[width=8cm]{Fig13.eps}}
\caption{Behaviour of the parameter $n_0$ as a function of the number of particles
in the simulation.  
\label{fig_n0N}
} 
\eef 
While the form of this profile is observed to be very stable in our
set of simulations, the parameters $n_0$ and $r_0$ vary as shown in
Figs.~\ref{fig_r0N} and \ref{fig_n0N}. Thus a good fit is given
by the simple behaviours $r_0 \propto N^{-1/3}$ and 
$n_0 \propto N^{2}$. In Fig.~\ref{fig_dpN} we show  the density 
profiles for various simulations with different $N$ where
the axes have been rescaled using these behaviours.
\bef
\centerline{\includegraphics*[width=8cm]{Fig14.eps}}
\caption{Density profile of the virialized structure at a time $t
  \approx 4 \tau_{scm}$ for simulations with different number of
  particles.  The y-axis has been normalized by $N^2$ and the x-axis by
  $N^{-1/3}$ (see text for explanations).  The behaviour of
  Eq.~(\ref{dpscm1}) is shown for comparison. 
\label{fig_dpN}
}
\eef

These  behaviours can be easily related to those we have 
observed in the previous section for $f^p$ and $K^p$. Indeed
using the ansatz of Eq.~(\ref{dpscm1}) it is straighforward to
calculate both the number of particles $N^n=(1-f^p)N$ which
are bound, and the binding energy $W^n$ of these 
particles [which is then
related to $K^p$ through Eq.~(\ref{energyapprox3})].
The first may be determined analytically as
\be
N^n=\sqrt{2}\pi^2 n_0 r_0^3
\ee
while the second may be written as
\be
W^n = -A Gm^2 n_0^2 r_0^5
\ee
where $A \approx 11$ is a numerically determined 
constant\footnote{For an isolated spherical system with a mass 
density profile $\rho(r)$ the potential energy is
$- 4 \pi G \int_0^\infty M(r) \rho(r) r dr$, where $M(r)$
is the mass enclosed in the radius $r$.}. 

The fitted behaviours for $r_0$ and $n_0$ thus correspond,
since $m\propto 1/N$, to $N^n \sim N$ (i.e. a constant 
bound mass, and therefore a constant ejected fraction
of the mass $f^p$) and $W^n \sim N^{1/3}$ (and therefore
$K^p \propto N^{1/3}$). Modifications of these fits 
consistent with the very slow variation of $f^p$
described in the previous subsection can be given.
Thus, to a good approximation, we simply find that 
the characteristic size of the final structure scales
with $N$ as $R_{min}$, the minimal radius attained in 
the collapse, does, i.e., in accordance with the
simple scaling argument which gives the prediction 
Eq.~(\ref{collapse-scaling}). Correspondingly the 
potential energy  which is bound in this 
structure, given that the mass is approximately 
constant, increases in proportion 
to the inverse of this scale (and scales also as
minimal value of the potential energy reached during
the collapse). 

\section{Mechanism of mass and energy ejection}
\label{Mechanism of mass and energy ejection}

In this section we consider in detail how the mass is ejected
from the collapsing system.  

\subsection{The ejected particles} 

An evident starting point in considering the ejection of mass
is to ask whether there is a direct correlation between a 
particle's initial radial position and the likelihood of its 
ejection, i.e., are particles preferentially ejected from the 
outside or the inside of the initial sphere? 

\begin{figure*}
\includegraphics[width=0.32\textwidth]{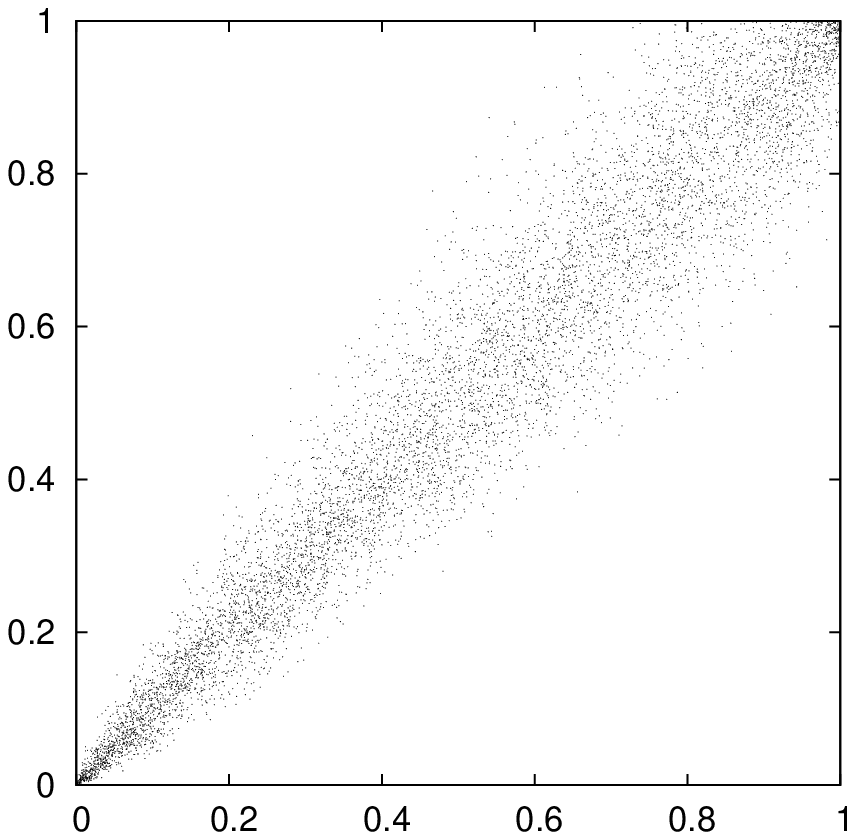}
\includegraphics[width=0.32\textwidth]{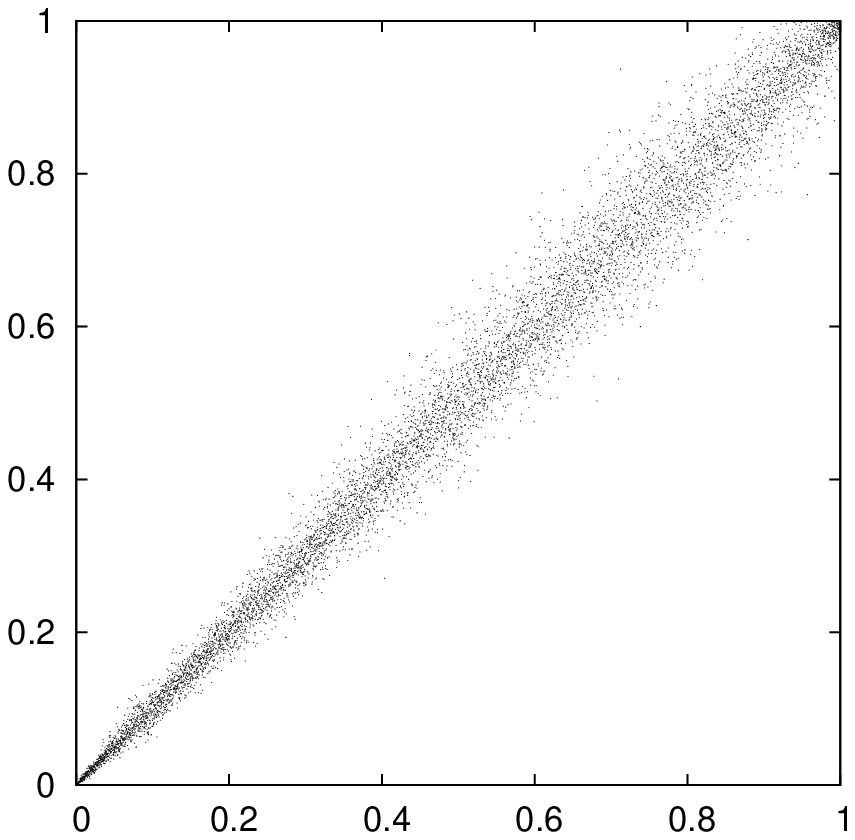}
\includegraphics[width=0.32\textwidth]{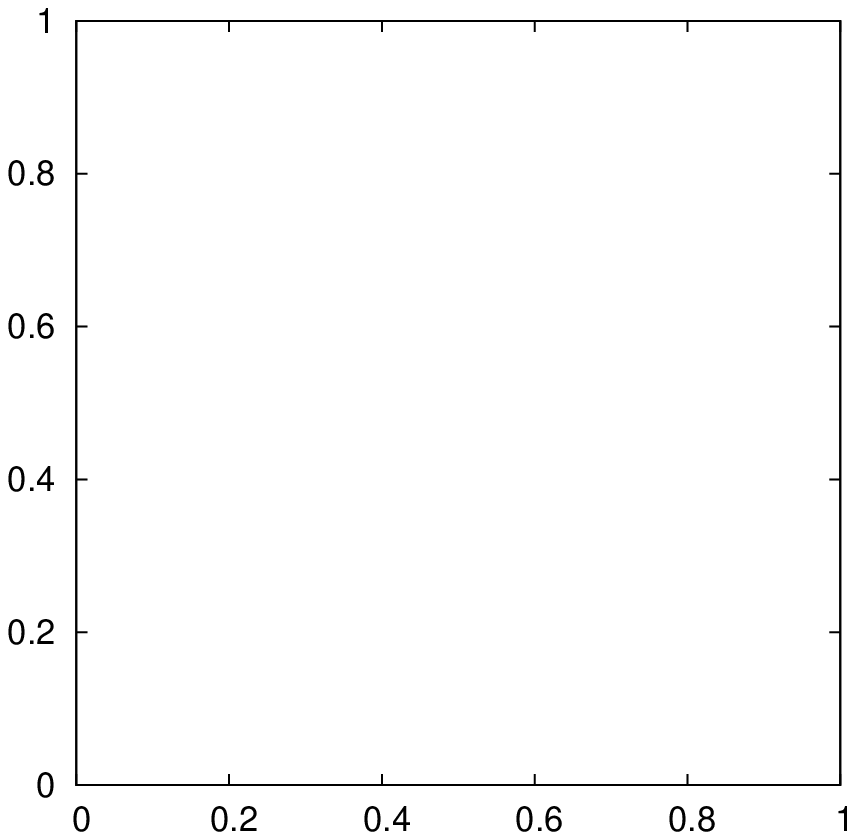}\\
\includegraphics[width=0.32\textwidth]{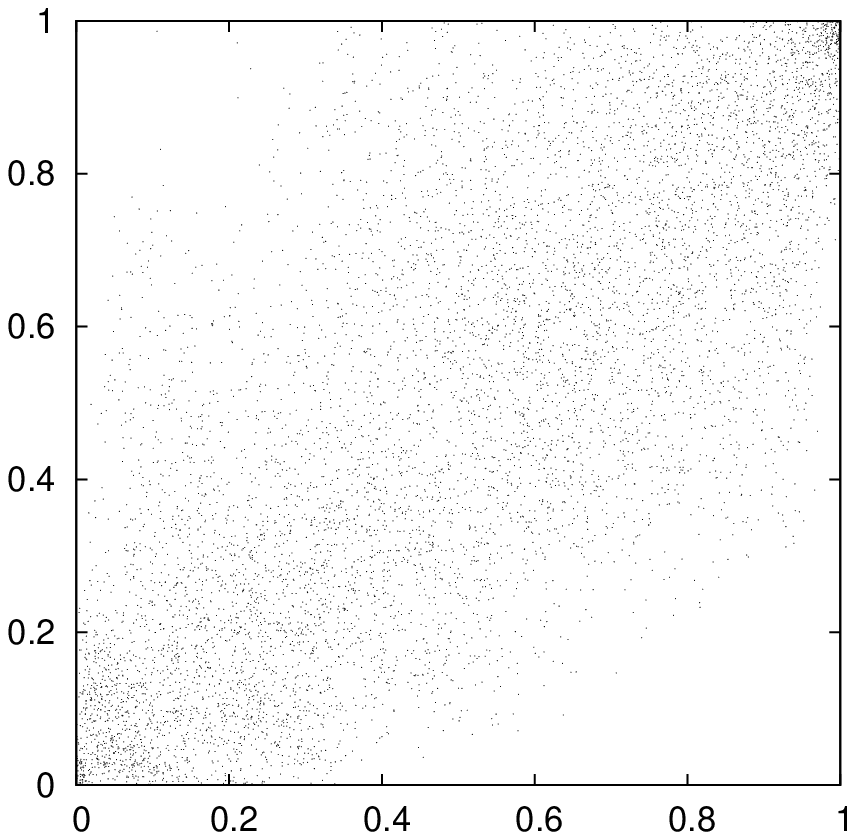}
\includegraphics[width=0.32\textwidth]{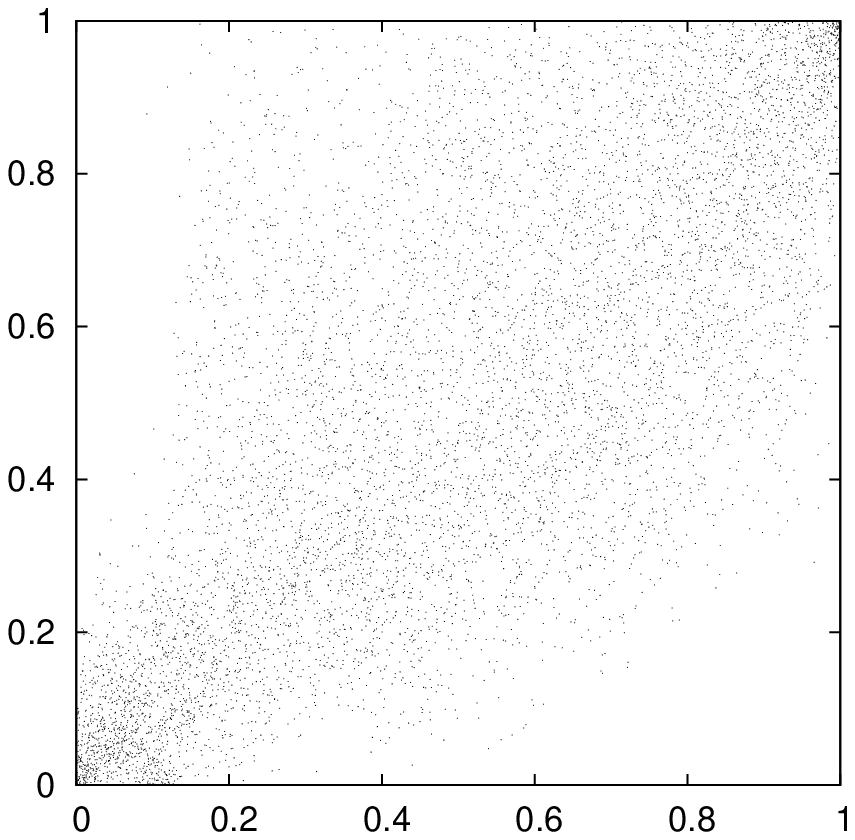}
\includegraphics[width=0.32\textwidth]{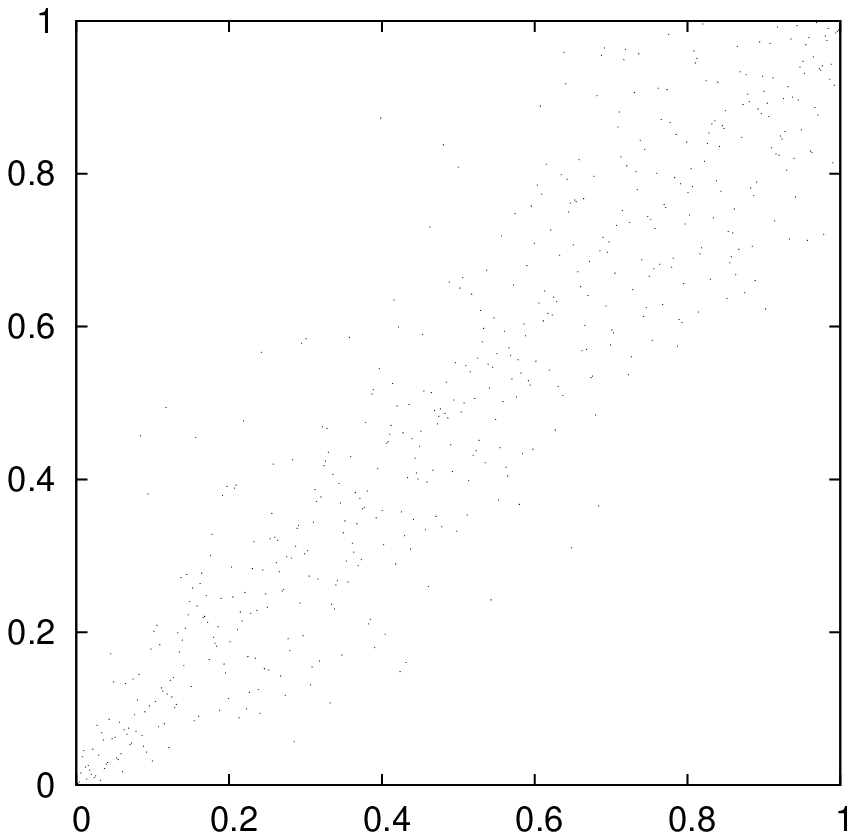}\\
\includegraphics[width=0.32\textwidth]{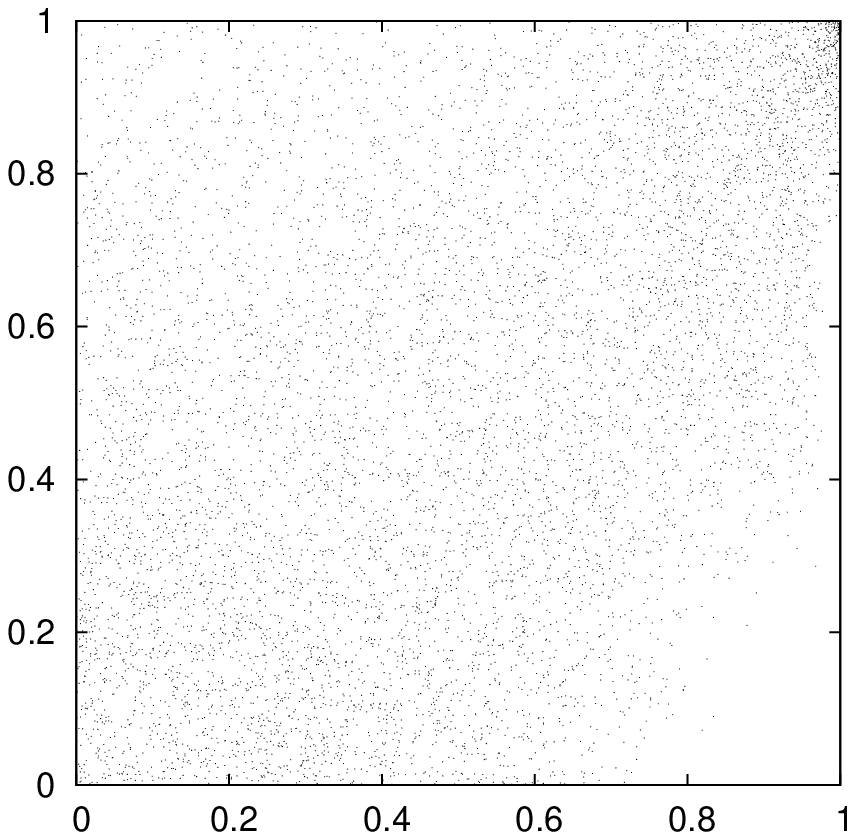}
\includegraphics[width=0.32\textwidth]{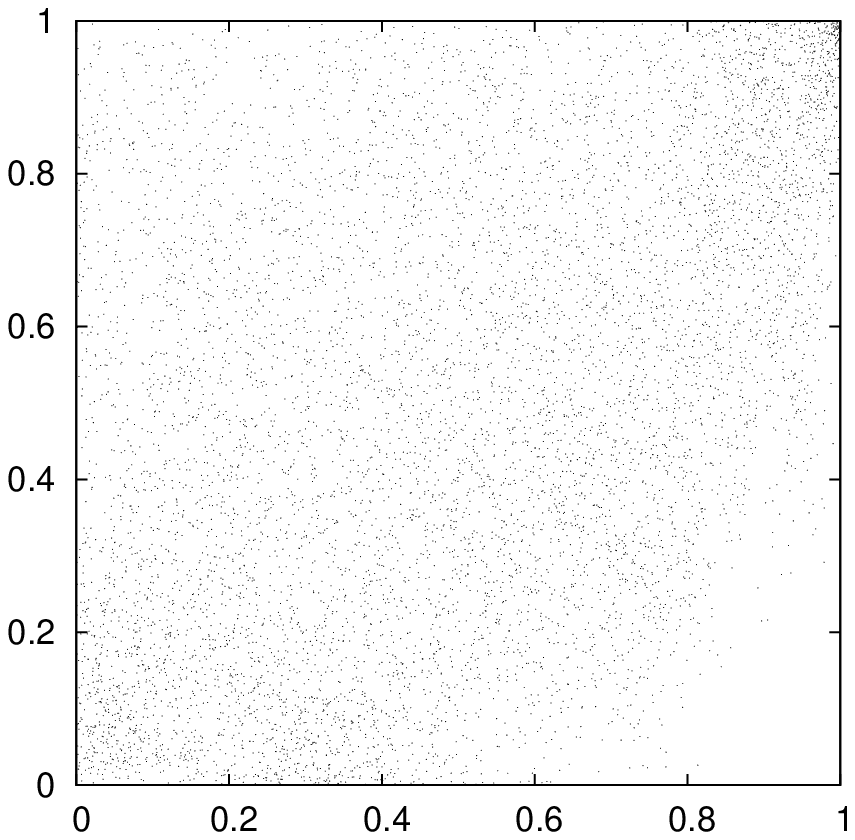}
\includegraphics[width=0.32\textwidth]{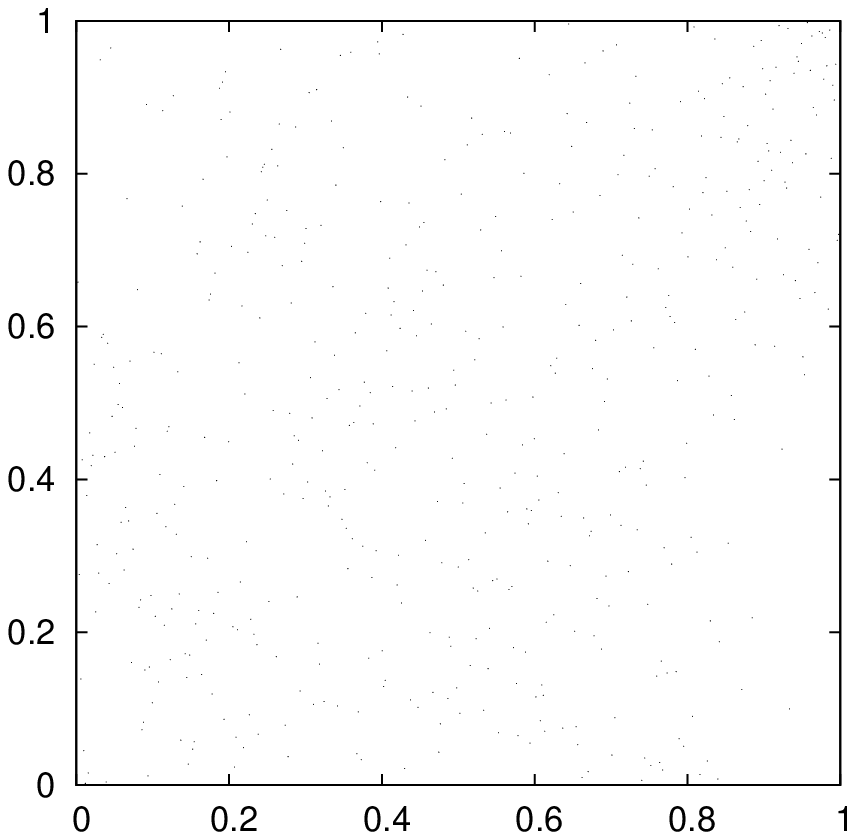}\\
\includegraphics[width=0.32\textwidth]{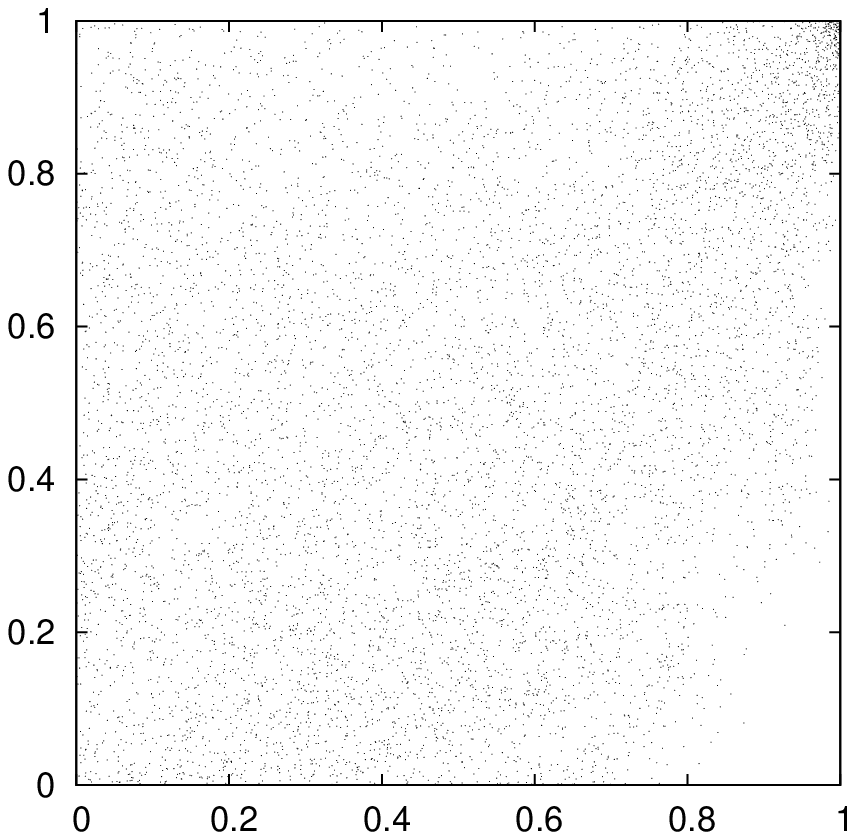}
\includegraphics[width=0.32\textwidth]{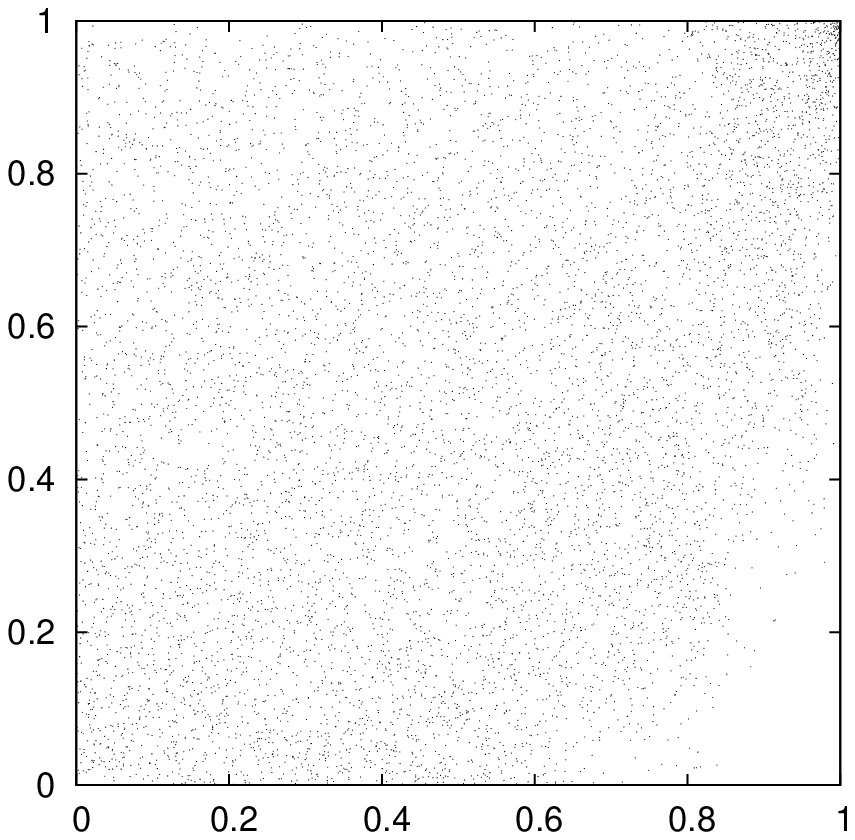}
\includegraphics[width=0.32\textwidth]{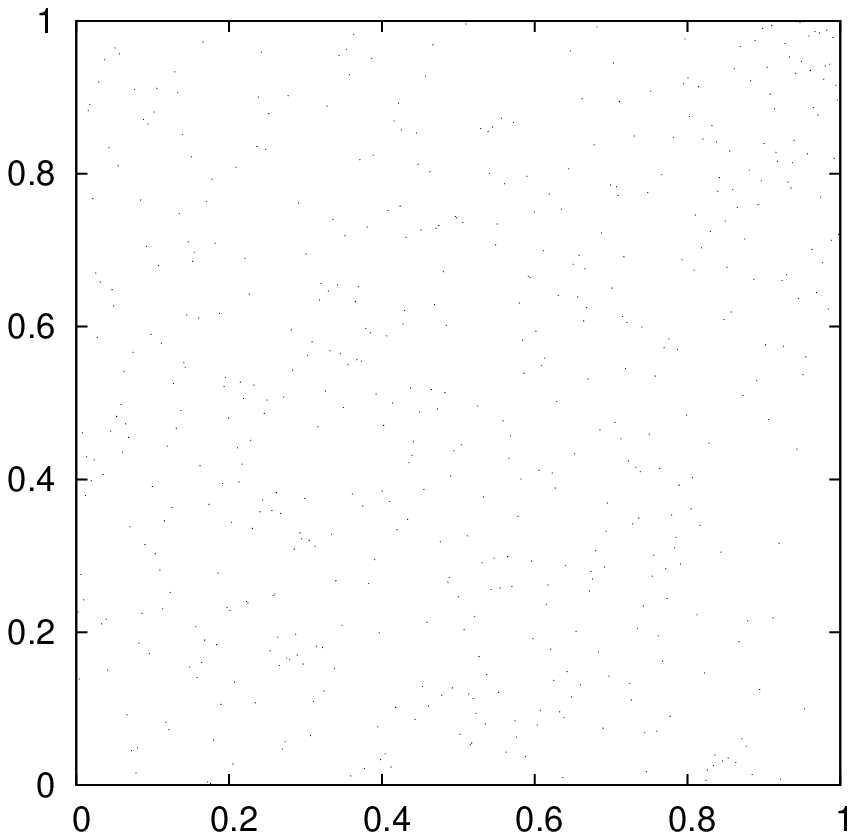}
\caption{Radial initial rank (horizontal axis) versus rank at the 
times indicated in Table \ref{table-times} for, from left to right, $N=131072$, $N=8192$ and $N=512$ particles and, from top to bottom, the times $t_1$, $t_2$, $t_3$ and $t_4$. For the plots corresponding of the system with $N=131072$ particles we have performed a random selection of $8076$ particles.
 \label{rank}}
\end{figure*}

Shown in Fig.~\ref{rank} is the ``radial rank 
correlation''~\footnote{This statistic has been introduced and used 
by \cite{aarseth_etal_1988} to examine 
the degree of correlation of particles' radial positions 
with their initial radial positions. From their data for
$N=5000$ it is concluded that such correlation
is very weak. }:
at the indicated time $t$ one ascribes a rank $n(t) \in [1,N]$ to 
each of the $N$ particles according to their radial distance from 
the centre of mass, and then for each point one plots 
the couple ($n(t)/N$, $n(0)/N$). The chosen times 
$t_1...t_4$ are not the same for the different $N$,
but are as given in Table~\ref{table-times}. 
\begin{table}
\begin{tabular}{@{}ccccc}
\hline
$N$ & $t_1$ & $t_2$ & $t_3$ & $t_4$  \\
\hline
$131072$ & $0.904$ & $0.988$ & $0.998$ & $0.999$ \\
$8192$ & $0.666$ & $0.952$ & $0.988$ & $0.998$ \\
$512$ & --- & $0.809$ & $0.953$ & $0.993$ \\
\hline
\end{tabular}
\caption{Times chosen for the different number of particles. 
\label{table-times}}
\end{table} 
This choice of the times is that discussed in 
Sect.~\ref{Predictions for scalings} above, i.e.,
it corresponds to comparing the different simulations
at times at which they are equivalent in the approximation
--- which leads, as explained in Sect.~\ref{Predictions for scalings}, 
to the predicted observed scaling of $R_{min}$ --- that 
they behave like infinite 
continuous systems with fluctations initially of 
amplitude proportional to $1/\sqrt{N}$, i.e., the times $t^\prime$ for the
particle number $N^\prime > N$ are determined
by 
\be
f[R(t^\prime)]\frac{1}{\sqrt{N^\prime}} =
f[R(t)]\frac{1}{\sqrt{N}} 
\ee
where $f(R)$ is the growth factor for perturbations, of
which the full expression is given in Eq.~(\ref{fullsolution}).

From these plots we see that that, while the strong
correlation evidently present at early times tends to disappear,
there is still some visible structure in the plot even at the 
final time, which corresponds to a scale factor just slightly
larger than $R_{min}$, i.e., just a very short time before 
the maximal collapse. 
We do not show the plots for later times --- after the
collapse --- as they are essentially identical to this last one.
Given that typically of order twenty per cent of the mass is 
ejected, it thus appears that although ejected mass originates
from all parts of the initial structure, it comes very 
preferentially from the outer regions of the sphere. Indeed the 
residual correlation in the plot appears to be a result 
solely of the correlation between ejection and initial radial 
position. 

The degree of this correlation can be seen much more clearly
in Fig.~\ref{fig_prt0}, which shows the estimated probability 
density function of ejection as a function of initial radius,
i.e., the normalized histogram of the ejected particles as
a function of their initial radius.
\bef
\centerline{\includegraphics*[width=8cm]{Fig16.eps}}
\caption{Normalized histogram (i.e. estimated probability
distribution function) of the ejected particles as a
function of their radial position in the initial configuration. 
The dashed line shows the best power-law fit.
\label{fig_prt0}} 
\eef 
In all the simulations we observe a very consistent 
behaviour, fitted approximately by a simple power-law: 
\be p_e(r) \propto r^4\,. 
\ee 
Given the uniform initial distribution, the probability that
a particle is at radius $r$ initially is proportional to
$r^2$, so this result corresponds to a conditional probability 
of ejection
\be p(r) \propto r^2 \,,
\ee 
i.e., the probability that a randomly chosen particle at radius $r$
in the initial configuration will be ejected grows approximately
as $r^2$.  Thus, although some of the ejected mass comes from 
the inner parts of the initial sphere, there a very clear 
systematic correlation between initial radial position and 
ejection.

\subsection{Development of lag during collapse}

Since particles are preferentially ejected from the initially
outer parts of the sphere, one would expect that their ejection 
may be related to a dependence of the evolution of particles 
prior to the collapse on their radial position. Indeed we 
have noted that the radial rank correlation plot of 
Fig.~\ref{rank} at the last time shown, just before the
collapse, is almost indistinguishable from that after
the collapse. As we have discussed, however, in the uniform 
limit of the spherical collapse model (SCM), ``particles'' do not 
``see'' the finite size of the system during the collapse phase,
and indeed perturbations about the SCM in the 
continuum (fluid) limit may be treated in this same
approximation that the collapsing system is infinite. With
these approximations therefore the evolution of the trajectory
of a particle cannot depend non-trivially on its 
initial radial position. As a corrollary such a dependence 
on radial position must arise necessarily from a coupling 
between the evolution of perturbations and the finite size 
of the system. 

To quantify the difference in evolution of particles as a function
of their initial radial position we consider the ``lag'', which
we estimate as follows:
\be
\label{lag-def}
\Lambda(r,t)=\frac{1}{N(r)}\sum_{r\leq|\br'|\leq r+\delta r}
\left( r'(t)-r'_{\rm SCM}(t) \right), 
\ee
where the sum is over the particles $N(r)$ initially at radial distance
between $r$ and $r+\delta r$ (with respect to the centre of mass), 
$r'(t)$ is actual radial distance at time $t$ of the particle
and $r'_{\rm SCM}(t)$ is the radial distance predicted at 
time $t$ in the SCM, i.e., $r'_{\rm SCM}(t)=R(t) r' (0)$.
The lag thus represents the average discrepancy between
a particle's radial position and that in the SCM, as a 
function of radius. 

\begin{figure*}
\psfrag{X}[c]{$r_0$}
\psfrag{Y}[c]{$\Lambda(t)/r_{\rm SCM\left(t_{N_i}\right)}$}
\includegraphics[width=0.45\textwidth]{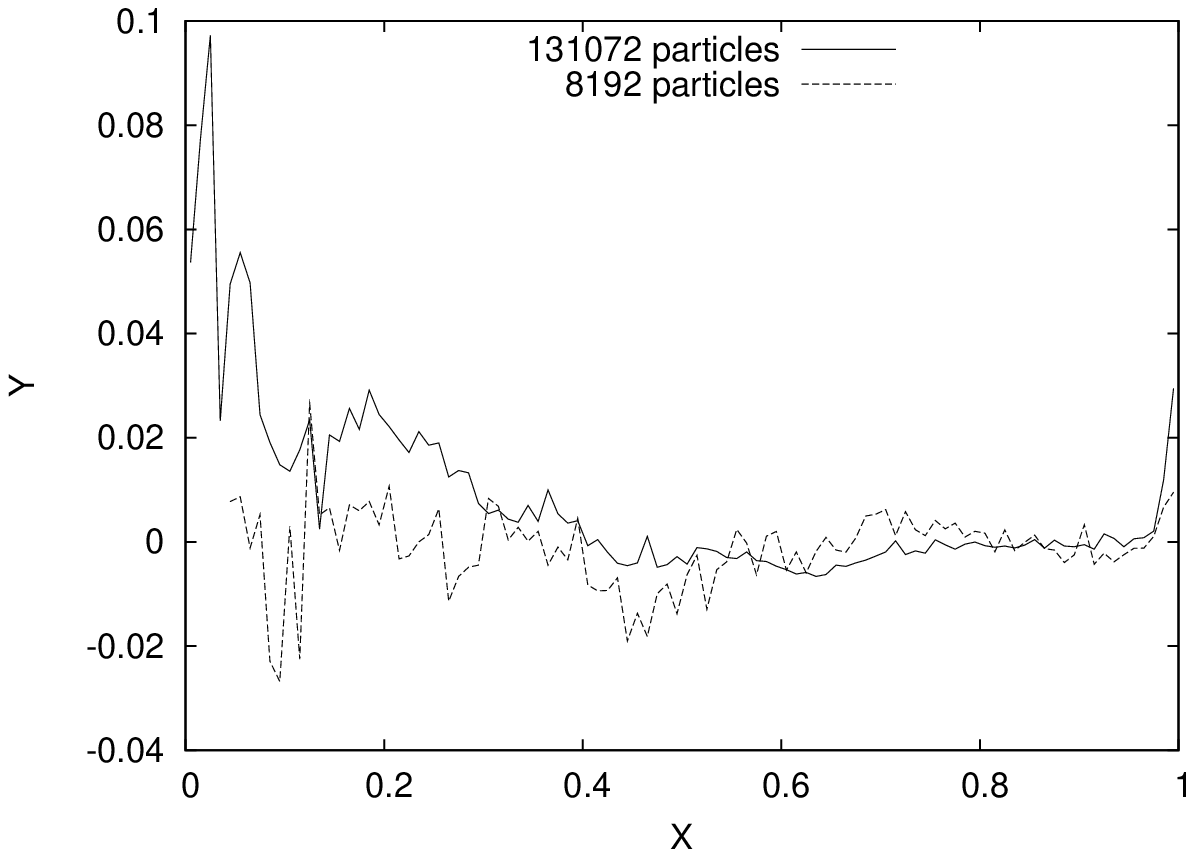}
\includegraphics[width=0.45\textwidth]{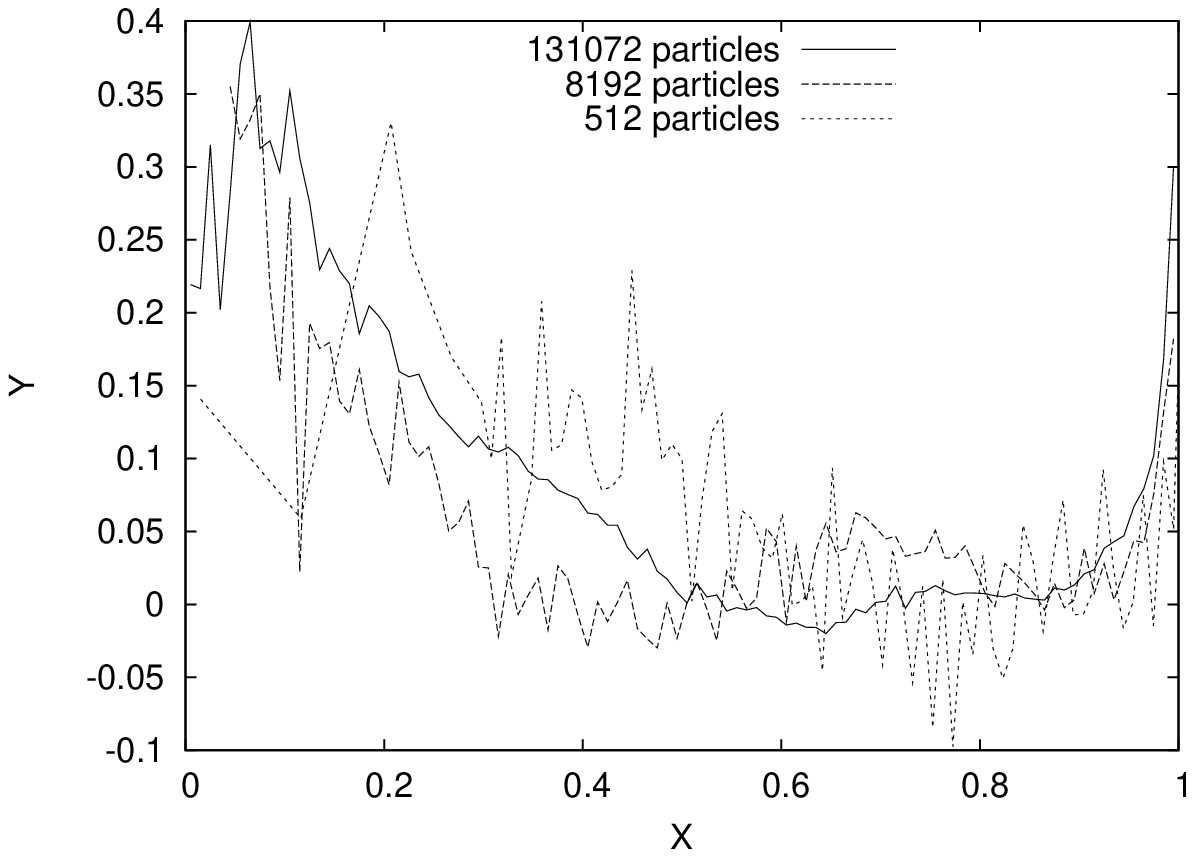}\\
\includegraphics[width=0.45\textwidth]{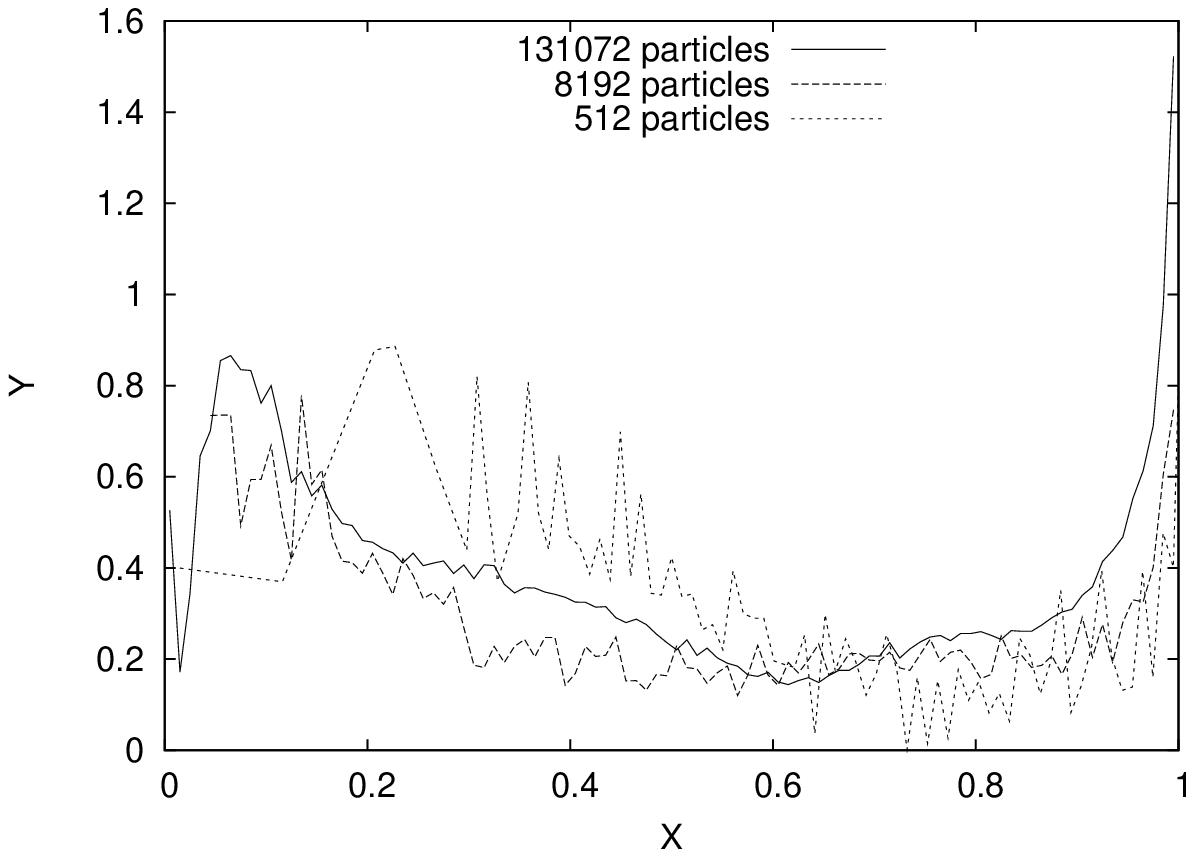}
\includegraphics[width=0.45\textwidth]{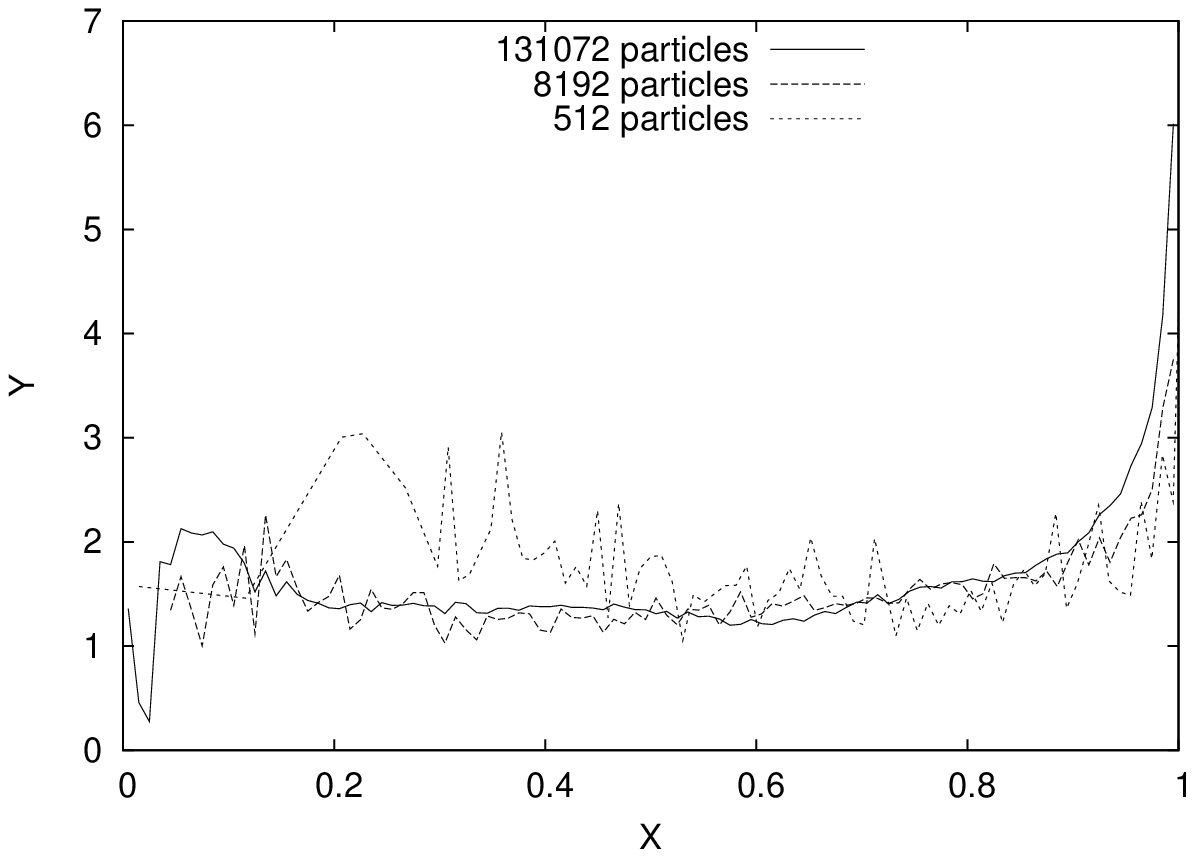}
\caption{The ``lag'' as defined in Eq.~(\ref{lag-def}), normalized
as indicated, for the same four times as in the previous figure. 
 \label{lag-fig}}
\end{figure*}

The results for this quantity for
different simulations and times --- the same as those in 
the previous figure of the radial rank correlation --- are
shown in Fig.~\ref{lag-fig}. For convenience of comparison
$\Lambda (r,t)$ has been normalized to the radius 
$r_{\rm SCM}(t)$ of the sphere in the SCM model at the 
corresponding time $t$. We see that a net positive lag
(i.e. net ``stretching'' of the positions with respect
to their SCM values) develops initially for two quite
separate ranges: for particles initially close to the 
centre, and for particles close to the outer 
boundary\footnote{Note that, since the mass inside the radius $r$ grows 
in proportion to $r^3$, the fraction of the mass lagging
in the two ranges is actually comparable even at the 
initial time despite the difference in radial range.}.
The reason for the development of systematically positive 
values is quite different in the two cases. For the particles
close to the centre there is a systematic effect which
makes the estimator necessarily biased towards positive
values: once particles can deviate from 
their SCM radius by of order $\Delta r$, particles within
this distance of the centre can reach the centre of 
mass, i.e., the minimal possible value of the radius,
and thus their radius will start to increase again,
which leads to an intrinsic asymmetry towards 
net positive values of $\Lambda(r,t)$ at these
scales. From the corresponding plot in Fig.~\ref{rank} 
we see that particles' radial positions with respect to the SCM
indeed grow in the course of the collapse, with
considerable redistribution of mass over the
entire system already at the third time slice.

The clear positive lag which develops first close to
the outer boundary, and then propagates progressively
into the volume as time goes on, can be understood as
follows. In addition to the mean field responsable for
the SCM motion, the particles move under the effect
of the field due to fluctuations which modify the 
SCM trajectories. These latter contributions are local,
with the dominant contribution coming initially from
nearest neighbour particles. For a shell at the outside
of the volume, there is thus a difference compared to one
in the bulk: as particles move around there is 
no compensating inward flux at the boundary for the mass 
which moves out under the effect of perturbations. Thus the 
net density of the outer shell decreases, and also the
average density in the sphere at the corresponding radius,
slowing its fall towards the origin\footnote{We note 
again, as we did in the introduction, that the
formation of a lagging low density outer region
(``halo'' ) during collaspe is noted 
in \cite{aarseth_etal_1988}, and a similar qualitative 
explanation for it is briefly outlined.}. As time
goes on this asymmetry propagates into the volume:
because the outer shells fall inward more slowly  the
mass in these shells stretches out radially, lowering
the density further, which in turn ``feeds'' less flux 
into the  shell. In the subsequent times in the figure
we see clearly this effect, with a larger part of the
outer mass lagging more and more compared to the SCM
model, and, more significantly, compared to the mass
closer to the centre. Indeed at the last time shown,
just before the minimal collapse radius is reached,  
the lag is essentially constant --- equivalent to 
a uniform dilatation of the whole sphere with respect
to the SCM --- for particles 
initially at radii less than approximately $0.8$,
but then rises sharply. The conclusion is that particles 
in the outer shells arrive at the centre of mass on 
average much later than those in the bulk.

\subsection{Mechanism of mass ejection}

\begin{figure}
\psfrag{X}[c]{$t$}
\psfrag{Y}[c]{Arbitrary units}
\includegraphics[width=0.45\textwidth]{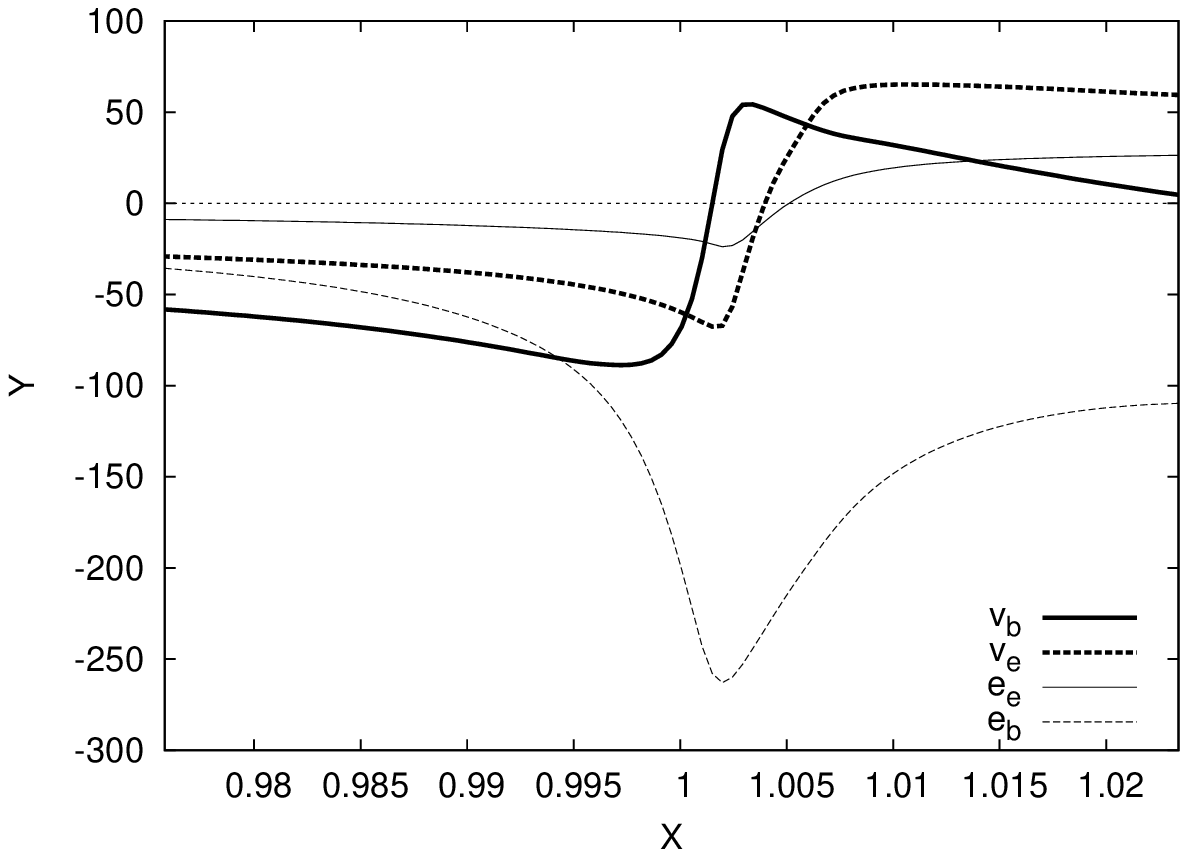}
\caption{Radial velocity, and average energy per particle, as a function
of time, of particles which are bound/ejected  at the end of 
a $P131072$ simulation. The energy of the particles 
has been arbitrarily rescaled.
 \label{radial-131072}}
\end{figure}

That it is this  ``late arrival'' of the outer parts
of the sphere which leads to their net gain in energy
and subsequent ejection can be seen from Fig.~\ref{radial-131072}.
It shows, for the P$131072$ simulation, the temporal
evolution of the components of the mass which are asymptotically
ejected or bound (i.e. with positive or negative energy a short
time after the collapse has occurred). More specifically 
it shows the evolution of $v_e$ (and $v_b$) which is
the average of the {\it radial} component of the velocity 
for the ejected (and bound) particles, and also e$_e$ (and e$_b$)
which is the mean energy per ejected (and bound) particle 
(i.e. the average  of the individual particle energies).
The behaviours of $v_e$ and $v_b$ show clearly that the
ejected particles are those which arrive on average late
at the centre of mass, with $v_e$ reaching its minimum
after the bound particles have started moving outward.
Considering the energies we see that it is in this 
short time, in which the former particles pass through 
the latter, that they pick up the additional energy
which leads to their ejection. Indeed the increase
of $e_e$ sets in just after the change in sign of $v_b$,
i.e., when the bound component has (on average) just 
``turned around'' and started moving outward again.
The mechanism of the gain of energy leading to ejection
is simply that the outer particles, arriving later on 
average, move through the time dependent {\it decreasing}
mean field potential produced by the re-expanding inner mass.
We note that a mechanism of this kind for ``mean-field'' 
mass ejection has been discussed in \cite{david+theuns_1989}
and \cite{theuns+david_1990}. Although we cannot quantitatively apply
the results of these latter works to the present case ---
they treat the case of oscillations about a quasi-equilibrium ---
they give a qualitative insight into the ejection we observe.

\bigskip

\subsection{Scaling of ejected energy}

Let us now finally attempt to be a little more quantitative, and 
relate this qualitative understanding of how the mass is ejected
to the actual scaling we have observed of the ejected energy as
a function of $N$.
 
To a first approximation the curves at the latest time plotted
in Fig.~\ref{lag-fig} overlap, i.e., the profile describing 
the lag of the outer shells with respect to the inner ones
is the same. In this approximation all the simulations
are therefore equivalent, up to rescalings of time and 
length, and would be expected to eject the same fraction
of their mass if this ejection arises from the delay described
by these curves. 

A simple estimate of the energy $E_g$ gained by the outer lagging
mass as it passes through the inner core is then 
\be
E_g \sim \frac{W_{min}} {t_{char}}\times \frac{R_{min}}{ v_{char}}
\ee
where the first term represents the rate of change of potential
produced by the ``core'' structure, and the second the typical 
time for particles to cross it, where the parameters $t_{char}$ and 
$v_{char}$ are characteristic time and velocity scale for the
process.
We have seen already numerically that, to a very good
approximation, $W_{min} \propto R_{min}^{-1}$, and
therefore we have simply 
\be
\label{estimate}
E_g  \sim \frac{1}{ t_{char}} \times \frac{1 }{v_{char}}\;.
\ee

We expect the role of $t_{char}$ to be played by the typical
time over which the ``fall through'' leading to the transfer 
of energy between the components occurs as we have described
above. This can be estimated numerically, and its scaling 
with $N$ inferred, from the data
we have shown. For example, estimating it from the curve 
for the evolution of the total kinetic energy, which reaches
a sharp peak as we go through this phase, as the temporal 
width at half the maximal value of the kinetic energy,
we find the results shown in Fig.~\ref{fig_dt}. As shown a
very good fit is given by the scaling $\Delta t \propto  N^{-1/2}$
\bef
\centerline{\includegraphics*[width=8cm]{Fig19.eps}}
\caption{ Time interval of the ``peak'' of the kinetic energy
as a function of $N$.
\label{fig_dt}
}
\eef

The scaling of this characteristic time coincides with that 
predicted by the perturbed SCM model, as described in
Sect.~\ref{Predictions for scalings}, for the 
difference $t_{max} - \tau_{scm}$, where $t_{max}$ is 
the time at which the maximal collapse is reached. That this
latter quantity also scales itself as predicted can be seen in
Fig.~\ref{fig_tmax}.
\bef
\centerline{\includegraphics*[width=8cm]{Fig20.eps}}
\caption{ Time $t=t_{max} - \tau_{scm}$ as a function of $N$. 
\label{fig_tmax}
}
\eef

\begin{figure}
\psfrag{X}[c]{$t$}
\psfrag{Y}[c]{$v_r$}
\centerline{\includegraphics[width=0.45\textwidth]{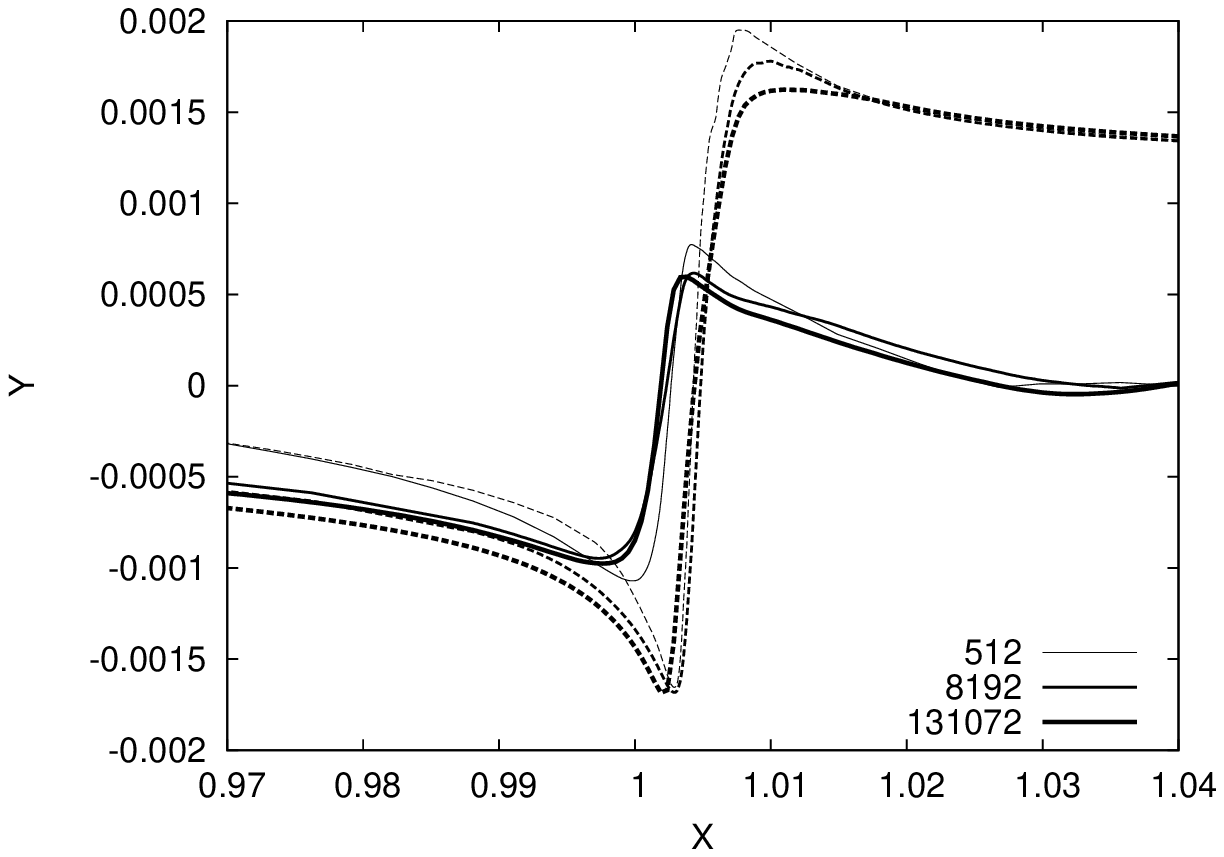}}
\caption{Radial velocity, as a function
of time, of particles which are bound (solid lines) or expelled 
(dashed lines) at the end of the different simulations indicated.
For the $8192$ and $131072$ simulations, the x-axis 
and y-axis have been rescaled as described in the text.
\label{radial-velocities-rescaled}}
\end{figure}

The characteristic velocity $v_{char}$ is that of the 
particles which will be ejected as they pass through the collapse,
i.e., $v_e$ in Fig.~\ref{radial-131072} above. We might 
anticipate that its scaling will also
be predicted by the SCM for the maximal velocity reached
at the collapse, [cf. Eq.~(\ref{vmax-scaling}) above],
i.e., 
\be
v_{char} \propto N^{1/6} \;.
\label{velocity-scaling}
\ee
That this is indeed a good approximation to the scaling
can be seen in Fig.~\ref{radial-velocities-rescaled}, which
shows the same two radial velocities $v_e$ and $v_b$ as
in the previous figure. Both axes of the plots 
for the curves from the $P512$ and
$P8192$ simulations have been rescaled: the time
axis as just described, and the velocities according 
to Eq.~(\ref{velocity-scaling}). We see that in these
rescaled variables the evolution is indeed very
similar.

Taking these scalings to be exact, Eq.~(\ref{estimate})
then gives
\be
E_g \sim N^{1/3} \,.
\ee

This simple estimate agrees quite well with the observed scaling
of the ejected energy (i.e. asymptotic kinetic energy $K^p$ of
the positive energy particles). We have seen, however, that
this scaling is followed more precisely by $K^p/f^p$, i.e.,
the energy ejected per unit mass, while $f^p$ has a very weak
(approximately logarithmic) dependence on $N$. The simple result 
obtained here has in fact been obtained with the assumption 
that the mass ejected is independent of $N$. Assuming the 
relevant properties for understanding the ejected energy are 
given by the curves for the lag shown in Fig.~\ref{lag-fig}, 
this corresponded to the approximation that the curves at
the latest time, just before the collapse, are the same 
for all $N$. In fact we can see that this is only true to
a first approximation, and specifically that the integrated
``lagging'' mass (i.e. under the curve ) indeed appears to grow slowly as a 
function of $N$. Thus the scaling result obtained above can
be adapted taking into account this increase as a function of
$N$ of the mass. We do not, however, have an analytical 
understanding of this observed $N$ dependence of the 
lagging (and subsequently ejected) mass. 



\section{The Vlasov Poisson limit and sensitivity to initial conditions}
\label{mean-field}

\subsection{Definition of the Vlasov Poisson limit}

We now discuss the extent to which the evolution
of the simulated system is ``collisionless'', i.e.,
described by the coupled Vlasov-Poisson (VP) equations
(or ``collisionless Boltzmann equation'').  The
latter should describe the evolution of the
system in an appropriate $N \rightarrow \infty$.
The naive such limit applied here, i.e. 
$N \rightarrow \infty$ Poisson distributed 
particles, is clearly not the desired limit, 
as it converges to the singular SCM. Indeed we have 
explicitly identified macroscopic $N$ dependences 
in various quantities,
which diverge if we perform this naive extrapolation.

Existing formal proofs of the validity of the 
VP limit \citep{braun+hepp} for a self-gravitating
system require, however, that the singularity in
the gravitational force at zero separation be
regulated when the limit $N \rightarrow \infty$
is taken\footnote{This 
proof is given in a finite volume. We do not believe 
that this is an important difference here, as the 
dynamics we describe will be modified in a
quite trivial way, on the time scales 
considered, if we put our system in a box: the ejected component will
simply bounce off the walls and remain as
an unbound cloud moving in and out of the 
time independent potential of the ``core''.}.
Incorporating such a modification of the
force will clearly regulate the divergence
in the SCM as we converge to the uniform mass 
distribution. One would then expect to obtain,
for sufficiently large $N$, a final state which 
is well defined and $N$ independent, but 
strongly dependent on the implementation
(and scale) of the regulation.

This limit also is not the VP limit relevant here: 
while we have indeed introduced such a regulation of the 
force (characterized by the smoothing $\varepsilon$),
we have done so, as discussed in Sect.~\ref{Basic results}, 
for reasons of numerical convenience. Indeed,
as we have discussed, our criterion for our 
choice of $\varepsilon$ is that it be sufficiently small 
so that our numerical results are independent of it,
and we interpret our results as being representative 
of the limit $\varepsilon=0$. In Fig.~\ref{fig_diffeps} are
shown, for example, the evolution of the fraction of 
particles  with positive energy as a function of time 
for the different indicated values of $\varepsilon$.
\bef
\centerline{\includegraphics*[width=8cm]{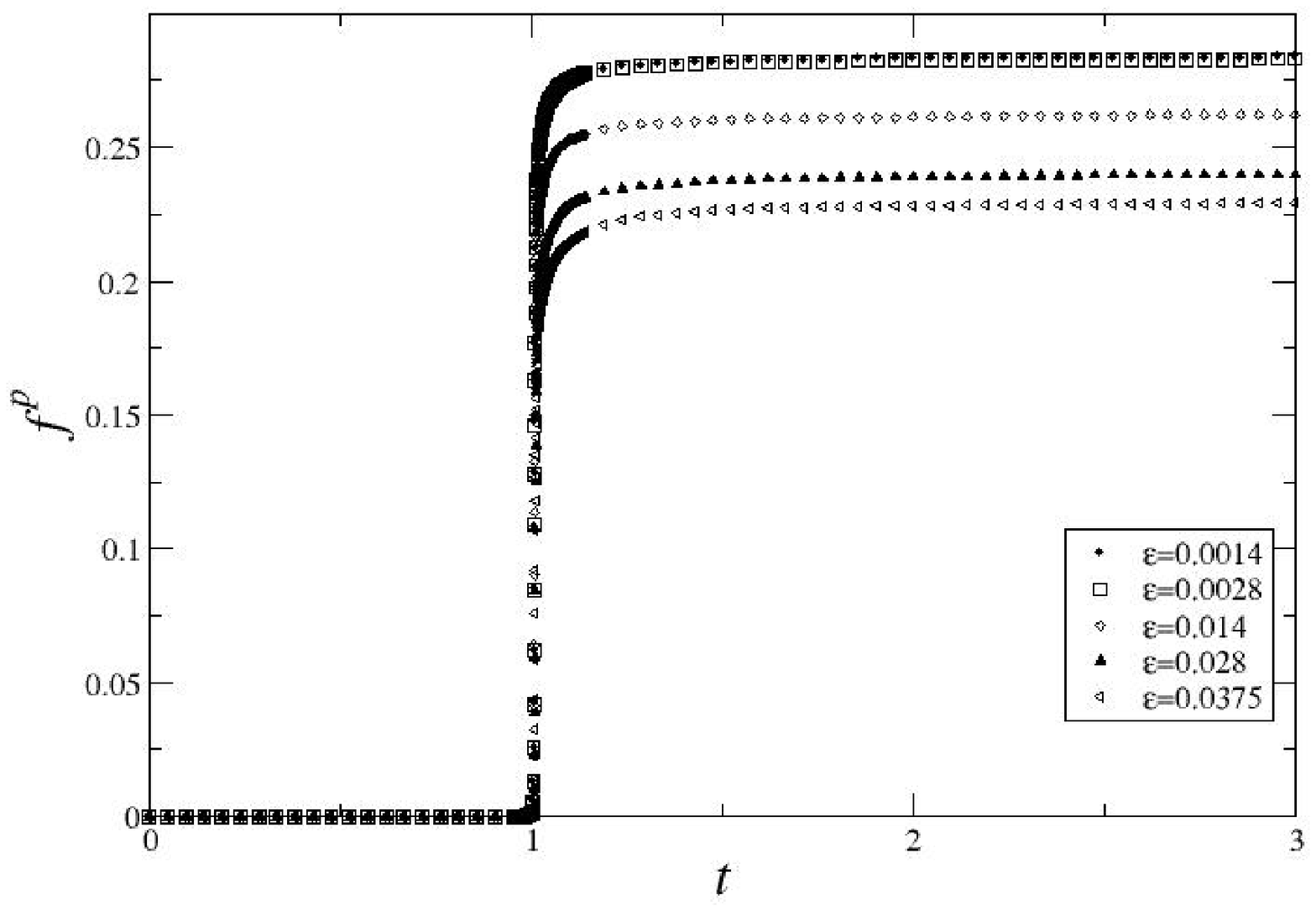}}
\caption{Evolution of the fraction of the mass with positive energy for
simulations with $N=32768$ for the different values indicated
of the smoothing parameter $\varepsilon$.
\label{fig_diffeps}
}
\eef
Other quantities we have considered show equally good convergence
as $\varepsilon$ decreases. Note that for the given simulation
(cf. Table~\ref{table-one}), the mean interparticle distance
$\ell=0.016$, so that the convergence of results is attained once
$\varepsilon$ is significantly less than $\ell$.  As we have 
seen, the minimal size reached by the collapsing system scales 
in proportion to
$\ell$ (with a numerical factor very close to unity, as can
be seen in Fig.~\ref{fig_rmin}).
We interpret the observed convergence as due to the 
fact that the evolution of the system is determined primarily
by fluctuations on length scales between this scale
and the size of the system. Once $\varepsilon$ is sufficiently 
small to resolve these length scales at all times, convergence
is obtained.

A different, less rigorous but more physically intuitive, derivation 
of the VP limit is given (see e.g. \cite{buchert_dominguez}) by a
coarse-graining of the exact one particle
distribution function over a window in
phase space. The VP equations are obtained for the coarse-grained
phase space density when terms describing perturbations in 
velocity and force below the scale of the coarse-graining 
are neglected. A system is thus well described by this 
continuum VP limit if the effects of fluctuations below 
some sufficiently small scale play no role in the evolution.
The validity of such an assumption for our system
is indicated by precisely the kind of behaviour we have 
just noted of our results as a function of $\varepsilon$. 

\subsection{Numerical extrapolation to the VP limit}

The VP limit defined in this way thus corresponds to
taking $N \rightarrow \infty$, while {\it keeping 
fixed the initial fluctuations above some scale}. 
The validity of this limit for our system may be explored 
numerically by defining such an extrapolation and studying
stability of our results to it. This can be done as
follows:  starting from a given Poissonian initial 
condition of $N$ particles in a sphere of radius $R$, 
we create a configuration with $N'=nN$ particles
by substituting each particle by $n$ particles 
in a cube of side $2r_{s}$
centred on the original particle. The latter
particles are distributed randomly in the
cube, with the additional constraint that their 
centre of mass is located at the
centre of the cube (i.e. the centre of mass 
is conserved by the ``splitting''). In this new 
point distribution
fluctuations on scales larger than $r_s$ are 
essentially unchanged compared to those in the 
original distribution, while fluctuations at and 
below this scale are modified\footnote{See \cite{gabrielli+joyce_2008}
for a detailed study of how fluctuations are 
modified
by such ``cloud processes''.}. We have 
performed this experiment for a Poisson initial
condition with $N=4096$ particles, splitting 
each particle into eight ($n=8$) to obtain
an initial condition with $N'=32768$ particles.
Results are shown in Fig.~\ref{fig_split} for 
the ejected mass as a function of time, for
a range of values of the parameter $r_s$,
expressed in terms of $\ell$, the mean interparticle
separation (in the original distribution).
While for $r_s=0.8\ell$ the curve of ejected particles 
is indistinguishable in the plot from the one for the 
original distribution, differences can
be seen for the other values, greater discrepancy
becoming evident as $r_s$ increases. This behaviour
is clearly consistent with the conjecture that the 
macroscopic evolution  of the system depends only 
on initial fluctuations above some scale, and
that this scale is of order the initial 
interparticle separation $\ell$. And, as anticipated,
this translates into an $N$ independence of the 
results when $N$ is extrapolated in this way for
an $r_s$ smaller than this scale.

The observed behaviour of the ejected mass for the larger
values of $r_s$ can be understood as follows: for an initial Poisson
distribution, after dividing each particles following the procedure
described above, the distribution remains Poissonian at large scale.
Therefore the growth of fluctuations inside the sphere and hence the
radius of minimal collapse, etc., will be the same for the original
and the new simulation. The relevant difference between the initial
conditions of the two simulations are for radius $r>R-r_s$ in the
original distribution, because particles situated in this region have
a strong probability to lie outside the original sphere in the
split distribution. These particles will lag strongly because they
feel a weaker effective gravitational field than the ones inside the
sphere and a large number of them will be ejected, explaining the
strong increase of ejected particles as $r_s$ increases.
\bef 
\psfrag{X}[c]{ \large$t$}
\psfrag{Y}[c]{ \large$f^p$}
\centerline{\includegraphics*[width=8cm]{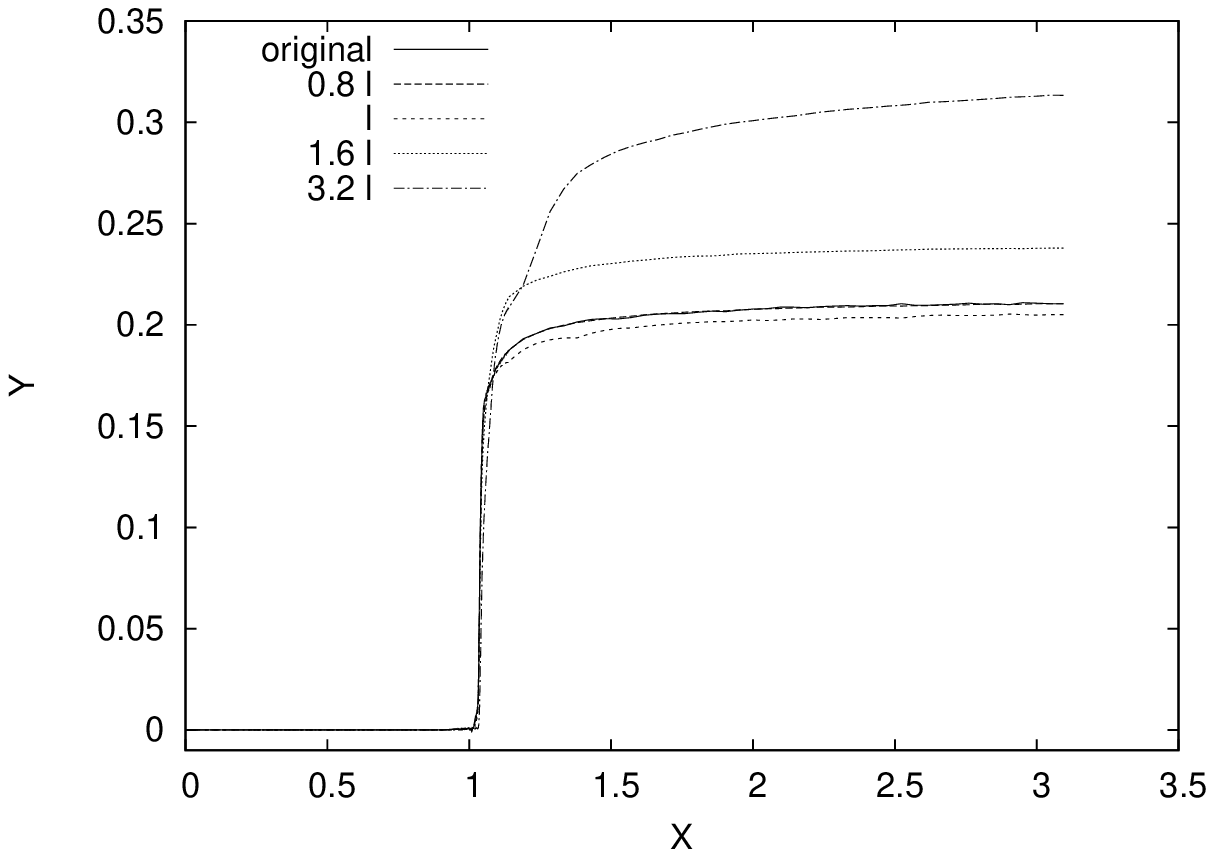}}
\caption{Evolution of the fraction of particles with positive energy 
as a function of time, for the different indicated values of the
parameter $r_s$ described in text. The ``original'' initial conditions
has $4096$ particles while the others have $32768$ particles. Note:
the curve for $0.8 \ell$ is not visible because it is superimposed
on that for the ``original'' one .
\label{fig_split}}
\eef

\subsection{Sensitivity to initial fluctuations at large scales}
\label{sensitivity}
\bef 
\psfrag{X}[c]{ \large$t$}
\psfrag{Y}[c]{ \large$f^p$}
\centerline{\includegraphics*[width=8cm]{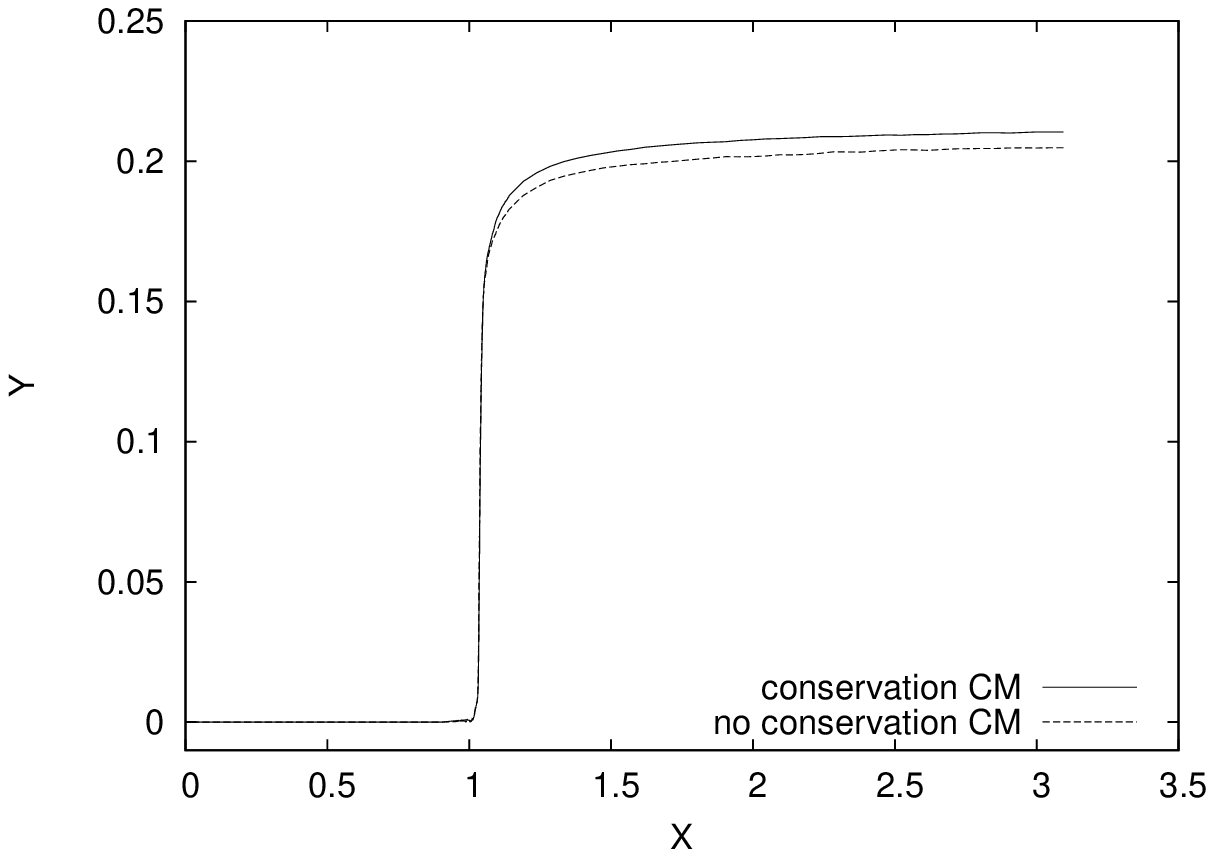}}
\caption{Evolution of the fraction of particles with positive energy 
as a function of time, from initial conditions given by the ``split''
configurations described in text. The results are for a Poisson 
configuration with $N=4096$ split into one with $N'=32768$ particles,
and $r_s=0.8 \ell$, and differ only in whether the centre of
mass of the split particles is conserved or not.
\label{fig_split_noCM}}
\eef

It is interesting to probe further the sensitivity of the results to 
the initial fluctuations over the range of scales which appear to be
relevant to the determination of the final state.  
In Fig.~\ref{fig_split_noCM} are shown the same quantity as in
the previous figure, for $r_s=0.8\ell$,  with the only difference 
that the $n$ points are now distributed {\it randomly} in a cube 
of side $2r_s$ both with and without the constraint that 
the centre of mass be conserved. Without the constraint 
there is an additional random displacement of the mass, which
induces slightly greater fluctuations at large scales, and
specifically a non-zero fluctuation in the dipole moment of the whole
mass distribution\footnote{For an infinite point distribution (see,
  e.g. ~\cite{book, gabrielli+joyce_2008}) the constrained case 
produces long wavelength fluctuations with a power spectrum 
decaying in amplitude in
  proportion to $k^4$ (where $k$ is wavenumber), while the
  unconstrained ``shuffling'' produces a power spectrum proportional
 to $k^2$.}.  We see that, although this corresponds to an extremely small
modification of the fluctuations at large scales, it leads
to a perceptible change in the final macroscopic properties.

Finally we can probe the dependence on initial fluctuations at
different scales by comparing the evolution from the Poissonian 
initial configuration, to that from initial  ``shuffled lattice'' 
configurations with the same number of particles. The latter
are generated by placing particles on a perfect lattice, and
then subjecting them, independently, to a random displacement 
in a cube of  characteristic size $\delta_s$. A sphere 
containing the required number of particles is then extracted,
with centre on a lattice point.
If the scale $\delta_s$ is or order the interparticle distance or larger, the
latter distribution has then, up to this scale, fluctuations which are
Poissonian (and identical in amplitude to those in the initial
Poissonian distribution), while at larger scales it has fluctuations
of a much lower amplitude than in the latter distribution. In
Figs.~\ref{fig_latt_mass} we show the evolution of the fraction of the
ejected mass as a function of time, for the indicated values
of the parameter $\delta_s$, for configurations with $4143$
particles. In comparison with the Poisson configurations 
we have reported above with approximately the same number
of particles (4096), which eject $20-25$ percent of their
mass, the mass ejected is significantly larger in all cases, increasing
monotonically as $\delta_s$ decreases. This is in line with 
what everything we have seen above: for any $\delta_s \leq 1$ 
the 
fluctuations at relevant scales are smaller than 
those in the Poisson distribution, and their amplitude
decreases as $\delta_s$ does.
For the limit case of a perfect lattice (i.e. $\delta_s=0$), 
the collapse is the most violent. Indeed in this case
the only density fluctuations regularizing the collapse
are surface fluctuations (associated with the finite size of
the system), and close to half ($43 \%$) of the initial
mass is ejected. 

\bef 
\psfrag{X}[c]{ \large$t$}
\psfrag{Y}[c]{ \large$f^p$}
\centerline{\includegraphics*[width=8cm]{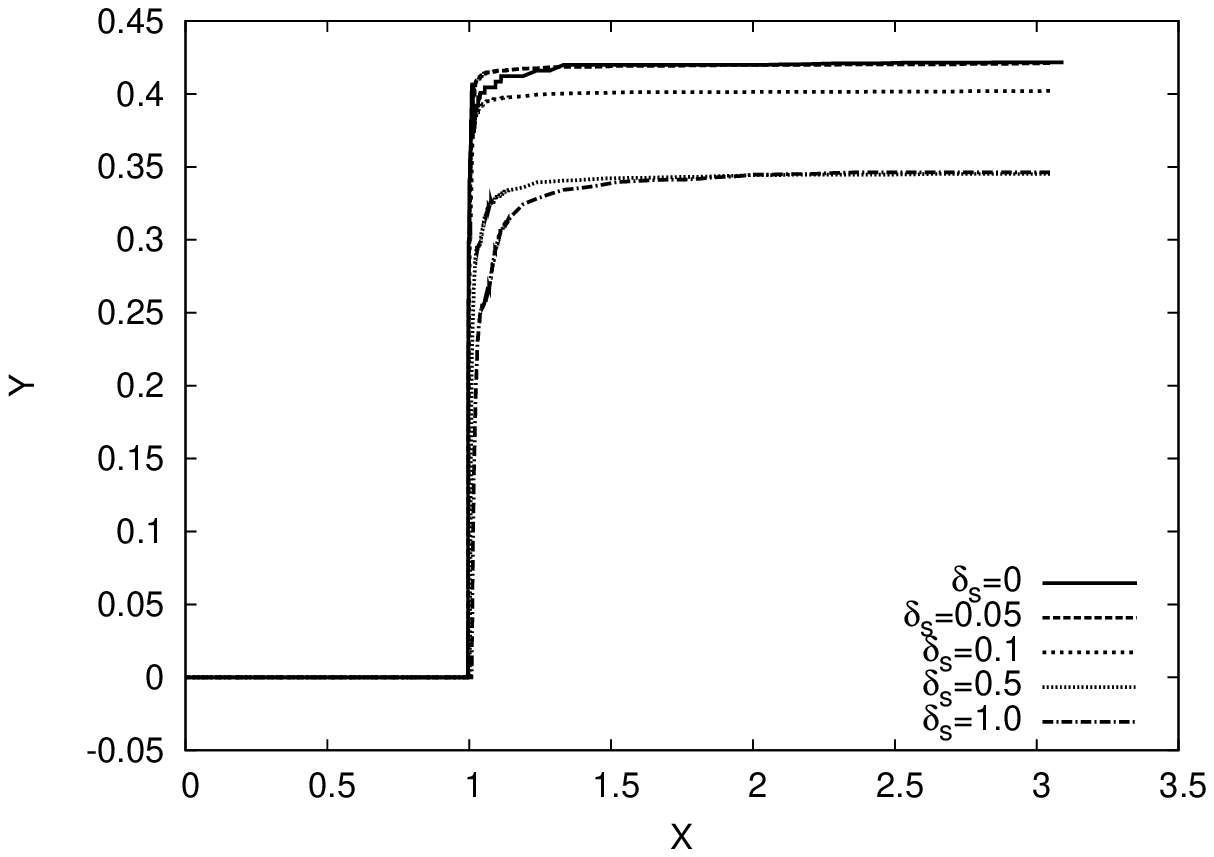}}
\caption{Evolution of the fraction of particles with positive energy 
as a function of time, for simulations from ``shuffled lattice''
initial conditions with different values of the parameter $\delta_s$.
  \label{fig_latt_mass}}
\eef

\subsection{Further test for non-VP effects}

\begin{figure}
\psfrag{X}[c]{$t$}
\psfrag{Y}[c]{$f^p$}
\includegraphics[width=0.45\textwidth]{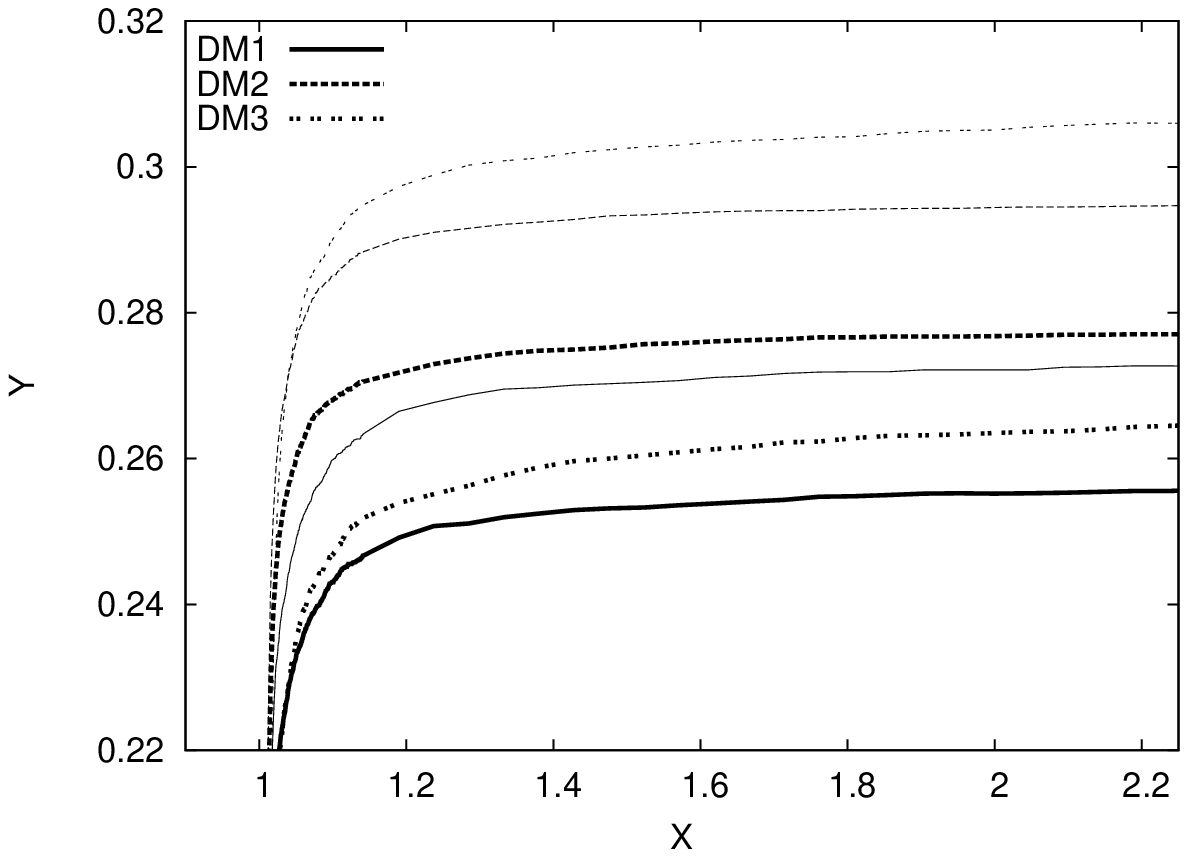}
\caption{Evolution of the fraction of particles with positive energy 
for the different indicated initial conditions (see text). In
each case the thick line corresponds to the heavier particles
(mass $m_1$) and the thin line to the lighter particles (mass $m_2$).
 \label{diffmass}}
\end{figure}

A useful test for non-VP effects is to use particles of different
masses in the simulations: given that the VP limit is essentially
a mean field approximation, the trajectories of particles should 
be independent of their mass if it is valid. There should therefore 
be no segregation (up to statistical fluctuations) between the 
different species in the final configuration if this limit applies. 
We have thus run three simulations with dispersion in particles' masses: we 
denote by DM1 and DM2 two realizations of an $N=4096$ Poissonian 
initial condition, in which we have assigned, randomly, a mass 
$m_1=4/3 m$ to half of the particles, and $m_2=2/3 m$ to the 
other half. Finally, we denote by DM3 the same realization
as DM2 (i.e. with particles in the same positions), but now
with half of the particles with masses $m_1=20/11 m$ and the other
half $m_2=2/11 m$. In Fig.~\ref{diffmass} we show the fraction of
ejected particles for the three simulations. Firstly we observe
that the difference between the number of light ($m_2$) and 
heavy ($m_1$) particles is, albeit larger in DM3 than in
DM1 and DM2, in all three cases of order the dispersion in the 
average number ejected in different 
realizations (see also Fig.~\ref{fig_fpos}). The difference in 
the number ejected of each species in a given simulation (with
given density fluctuations) is, however, too large to be a 
purely statistical fluctuation, and indeed we see that there are 
systematically more light particles ejected than heavy ones,
with a clearly more pronounced effect as the mass difference
increases. This small excess of lighter ejected particles can naturally
be attributed to the ejection of some (small) part of the mass
by two-body collisions, rather than by the mean field mechanism 
described in the previous section: such collisions modify the 
energy per unit mass, which is the relevant quantity in the mean 
field limit, in a different way according to their mass. In
a collision between a light particle and a heavy particle,
notably, the trajectory of the former is deviated much
more than that of the latter, and therefore it is more 
likely to be ejected\footnote{Such mass segregation
effects have been observed also in \cite{aarseth_1974}
in the ejection by two body collisions which occurs 
as such systems relax on much longer time scales than 
those considered here.}.

\section{Conclusions and discussion}

We have studied in this article the collapse and violent relaxation
of $N$ (classical Newtonian) self-gravitating particles, initially 
at rest and distributed
randomly in a sphere. We have focussed on the characterisation
of the macroscopic properties of the quasi-stationary virialized
state which results, and in particular on the mass and energy of 
the bound and ejected components. We have identified unexpected 
non-trivial dependences of the latter on $N$, where the latter
varies over approximately three orders of magnitude (from a
few times $10^2$ to a few times $10^5$). We have studied in 
detail the evolution through the collapsing phase, identifying
the physical mechanism leading to this mass and energy ejection.
Further we have used this to give a simple analysis which allows 
one to explain the observed scaling with $N$ of the energy per unit 
ejected mass ($\propto N^{1/3}$). Finally we have clarified how
one can test the hypothesis that the evolution is representative
of the Vlasov-Poisson limit by an appropriate extrapolation of
the particle number. While our main convergence test show 
good stability of results, another one allows us to detect 
residual non-zero effects corresponding to deviation from this
limit.
 
One point which we emphasize is that we have {\it not} explained, even
qualitatively, the very slow (approximately logarithmic) growth 
with $N$ of the fraction of the mass ejected $f^p$. We note that 
this result is numerically less robust than the result for 
the energy: the effect is both very small over the range of $N$ 
simulated, and also shows greater fluctuations from realization to
realization than for the energy per unit mass. Indeed in
Fig.~\ref{fig_fpos} we see that if we omit the two points
with largest $N$, the results would be quite consistent
with a convergence of the ejected mass to a constant value.
As these largest $N$ simulations are those we have not 
been able to test for discreteness effects using the tests
we have discussed (which require extrapolation to still larger
$N$), we cannot exclude that the apparent continued growth
might be due to such effects. Indeed we note the growth of
the ejected mass with $N$ cannot be consistently extrapolated by
a logarithmic (or power law) dependence to arbitrarily large $N$, 
as it is a quantity which is bounded above (by the total mass).
The numerical results we have given, despite the
fact that we have considered a very large number of particles,
thus do not actually resolve the asymptotically large
$N$ dependence (or asymptotic constant value) of the ejected
mass, or indeed that of the ejected energy (or the associated
virialized state) is. Further numerical study, with larger simulations,
would be feasible for groups with greater numerical resources 
than ours, and might resolve the 
issue\footnote{\cite{boily_etal_2002} gives results for
a spherical collapse with $10^7$ particles (but does not report
the ejected mass and energy)}. It is natural to expect that
the slow growth in the mass will reach an upper bound when
the fraction of mass becomes of order the total mass. We
remark that a  $\log N$ behaviour, for example, could be 
explained if the quantity of mass which ``lags'' as we have 
described, and is then ejected, grows
exponentially in the course of the evolution from a fluctuation 
which is originally of order $1/N$ (or some power thereof).
One would expect the growth law to change, and flatten to a constant, 
when the fraction of lagging mass becomes of order one half. 

In relation to this last point it is interesting also to consider 
a little further the case
of an initial lattice-like configuration, which we discussed
briefly in Sect.~\ref{sensitivity}. These are, as we have
noted, much more uniform configurations in the range of
scales which are relevant to the macroscopic evolution, and
indeed we observed considerably larger mass ejection.
This corresponds, as one would expect, also to a more
violent collapse, with larger energy ejection.
We show, for example, 
in  Fig.~\ref{fig_latt_kin} the evolution of the
total kinetic energy for the different values
of $\delta_s$ (the ``shuffling'' of the lattice) 
indicated. The increasing violence of
the collapse is evident as $\delta_s$ decreases. 
We note that, for sufficiently
small $\delta_s$, the maximal collapse (corresponding to
the maximum of the kinetic energy) actually occurs
{\it before} the time predicted by the uniform
SCM. This is in contrast to the behaviour for all
the Poissonian initial conditions we have considered,
which in all cases gave a collapse time longer than
$\tau_{scm}$ (cf. Fig.~\ref{fig_tmax}). In the
latter case this behaviour can be understood 
in terms of the lagging mass we have discussed,
which leads to the system having a lower
effective mean density (and therefore a 
longer collapse time). That a different behaviour
occur for the exact (or near exact) lattice,
may be indicative that a quite different 
behaviour of the Poissonian system may be 
reached when $N$ is extrapolated sufficiently
far that, like in the exact lattice, a regime
is reached in which the dominant fluctuations
are those arising from the finite size of the system.
A naive extrapolation of the $N$ dependence of
$f^p$ for the Poissonian configurations to
the value observed for the lattice gives
$N \sim 10^7$. To resolve this point it would 
evidently be interesting to study the evolution
from these initial lattice or lattice-like
configurations, and indeed  other very uniform 
configurations such as those described 
in \cite{Ha06}, in more detail, and for 
larger particle number\footnote{As the cost of
numerical integration depends, essentially, on the
violence of the collapse, we reach our limit for
the exact lattice at a particle number of order 
that reported (N=4143).}.
\bef 
\psfrag{X}[c]{ \large$t$}
\psfrag{Y}[c]{ \large$K(t)/m$}
\centerline{\includegraphics*[width=8cm]{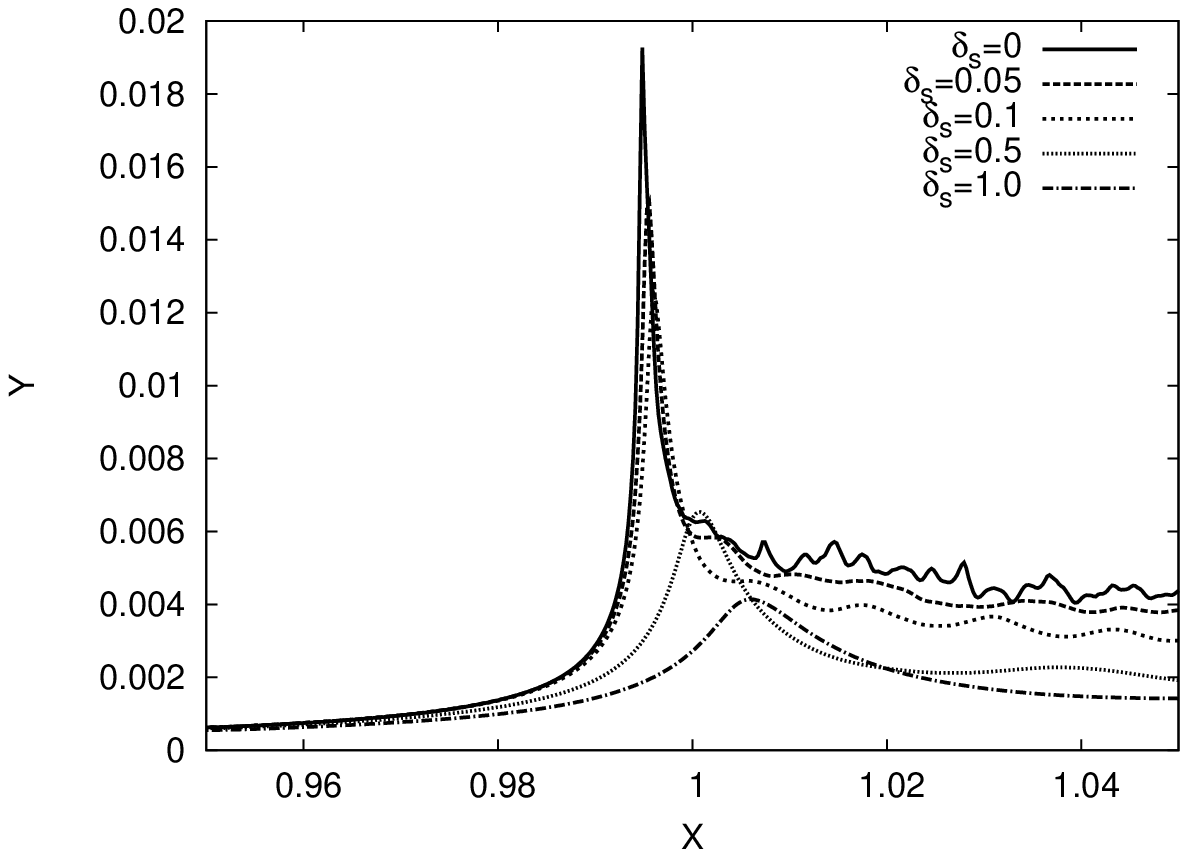}}
\caption{Evolution of the kinetic energy as a function of time,
for the shuffled lattice initial conditions with the indicated
values of $\delta_s$.
  \label{fig_latt_kin}}
\eef

What, otherwise, have we learnt from this study? The following points
seem pertinent to us:

\begin{itemize}

\item Analyses of violent relaxation usually neglect the effects
we have focussed on here, and assume that energy and mass 
of the final virialized system is equal to that of the initial
configuration. This is true, for example, of the use of the 
spherical collapse model in the context of 
cosmology (see e.g. \cite{halo}). While it may be, in this context,
a good approximation to neglect ejected mass --- generically we
would not expect to have the very large collapse factors 
observed here which are closely linked to this ejection ---
the non-trivial  slow $N$ dependences we have seen would urge some caution in 
this regard. Further in theoretical attempts to understand the
properties of the virialized states, the same approximation is
generically made. An evident example is the famous proposal by 
\cite{lyndenbell} that the result of violent relaxation should be
a state which maximizes a coarse-grained entropy in phase
space subject to the constraints of the VP equations, and
notably total energy and mass conservation. What we observe
in our simulations is that the density profile of the final
state (and also velocity distributions, which we have not
reported here) is very robust while the conservation of the
energy and mass of the initial state appears to be of little
relevance. This observation may be important in attempts to
understand the properties of these states, and in particular
their dependence (or non-dependence) on initial conditions.

\item Our study shows that the assumption that
the evolution of such systems on these timescales is fully
represented by the VP (or collisionless) limit should also
be treated with some caution. In our simulations we have
found small, but significant, non-VP effects with a simple
mass segregation test.
In any case it is important to define clearly the appropriate
extrapolation to the VP limit, which
we have seen may be less trivial than one might naively
appears. We believe the extrapolation we have defined here
--- in which particle number increases while the fluctuations 
above some scale are held fixed --- is
the appropriate one for this system. 

\item Our results suggest that open self-gravitating systems 
may be able, theoretically, to eject an arbitarily large amount 
of energy on a {\it dynamical} time scale. This effect is reminiscent of 
the ``gravothermal catastrophe'', and indeed its origin is
the same: because the gravitational potential is unbound
below, an arbitrarily large kinetic energy can be gained 
by making some mass arbitrarily bound. The mechanism, however,
and timescale, are completely different. It is clear that
the ejection we have seen is related to the collapse, which
in turn is related to the spherical symmetry of the problem
we have considered. What about more general initial 
conditions? We note that, although we expect smaller collapse
factors for a generic initial condition, very large collapse
factors may occur for non-symmetric initial conditions.
This question has been considered at length  
in \cite{boily_etal_2002}, which presents a detailed
numerical study of the generalisation of the fluid SCM 
to axisymmetric and triaxial configurations. In
the former singularities remain intact, and a
relation $R_{min} \propto N^{-1/6}$ is
found empirically to replace the $R_{min} \propto N^{-1/3}$ 
behaviour of the spherical case. In the triaxial case 
the collapse factors are found to be typically finite, 
but they can be very large and no upper bound is placed
on them.

\end{itemize}

The study we have presented is a purely theoretical one of
an idealized problem. It is interesting, of course,
to consider whether, in particular, the ``explosive'' ejection
of energy we have found could have any direct relevance 
to real physical systems or realistic models of them
in astrophysics or cosmology. For the former context 
it is relevant to consider the validity of the Newtonian
limit which we have treated. To do so it is useful
to write the ejected kinetic energy as
\begin{equation} 
K^p \sim N^{1/3}\frac{\Phi_I}{c^2} \times  M_I c^2
\end{equation} 
where $\Phi_I$ and $M_I$ are, respectively, the initial mass and
(Newtonian) gravitational potential, and  $c$ is the speed of light.
The first factor is, in fact, approximately the maximal value 
reached  by the potential, and if this is small compared to unity we
expect the Newtonian approximation to be valid. Thus our
determination of $K^p$ would be expected to remain valid 
until this energy reaches of order the evident upper bound
imposed by the rest energy of the ejected mass (which is 
of order that of the initial mass). We note also that
if the initial radius
of the sphere is $R_I$ the Newtonian approximation can
remain valid at all times provided the number of Poisson 
distributed masses satisfies
\begin{equation} 
N \ll \left( \frac{R_Ic^2}{GM_I} \right)^3\,.
\end{equation} 
Thus, for example, if $M_I$ and $R_I$ are taken to be a
solar mass and radius, respectively, the Newtonian
approximation should be valid at all times during the
collapse if $N < 10^{14}$, i.e., if the mass of the
initially Poisson distributed   ``particles''
is greater than about $10^{-14}$ solar masses. 
Normalizing the ejected energy using our simulations
above, one would obtain an ejected energy of
about $10^{49}$ ergs for the largest value of
$N$ we reported ($N \approx 2 \times 10^{5}$), and, extrapolating 
our results, an energy as large
as that of a typical supernovae ($\sim 10^{51}$ ergs)
for $N \sim 10^{11}$.
Whether such initial conditions (of very uniformly distributed
dark matter ``particles'') could possibly be produced 
in an astrophysical context and thus give rise to such purely
gravitational ``explosions'', with observational traces,
is beyond the scope of our study. In the context 
of cosmology we note that the results we have found may
be relevant, for example, in theories of structure formation 
in ``hot'' or ``warm'' dark matter cosmologies. In these
cases the initial spectrum of density fluctuations is cut-off 
abruptly below some cosmological scale, and the first 
structures should form at such very large scales, without
prior formation of structures at smaller scales.
In this context the physical effects we have observed might
then be expected to come into play, leading potentially also
to implications for the numerical resolution provided by 
an $N$-body discretisation.

Finally let us remark that rather than considering different 
non-uniform and/or non-spherical initial conditions, the next 
natural step in this study is to consider the effects of the 
presence of initial velocity dispersion. While we would
expect such dispersion to regulate the collapse and bound above
the ejected mass, it would be interesting to see the full
dependence of these quantities on $N$ and the amplitude of
this velocity dispersion. It would be interesting
to focus on a simple class of initial conditions for 
the velocities, e.g., ``waterbag'' distributions, 
studying very carefully the convergence of results 
for the virialized state as a function of $N$, and
to characterize carefully also the velocity 
distributions in the virialized states (which we
have not discussed here).
With respect to predictions of theoretical models such
as that of \cite{lyndenbell}, it would be interesting
to explore whether by appropriately modifying the 
energy/mass constraints on the final state better 
agreement may be obtained than has been observed
without such considerations (see, e.g., 
\cite{arad, levin_etal_2008}).

We thank the Centro E. Fermi (Rome) for the use of computing resources.
MJ thanks Steen Hansen for useful discussions, and the Istituto dei Sistemi
Complessi, CNR, Rome, for hospitality during several visits. 

\bibliographystyle{mn2e}

\appendix

\section{Growth of linearized perturbations}

After some simple algebra, Eq.~(\ref{deltaevol1})
reduces to 
\be
\label{deltaevol3}
2 \frac{d^2 \delta}{dR^2} \left( \frac{1}{R} -1 \right) +
 \frac{d \delta}{dR} \left( \frac{3}{R^2} -\frac{4}{R} \right) 
- \frac{3}{R^3} \delta =0,
\ee
where we have taken  $R(t=0)=1$. The general solution can 
be written as
\bea
&&
\delta(R) =  A_1 \frac{\sqrt{R-1}}{R^{3/2}} +
\\ \nonumber
&&
A_2 \frac{1}{R^{3/2}} \left(
\sqrt{R} (R-3) + 3 \sqrt{R-1}\;\;  \mbox{asinh}(\sqrt{R-1})  
\right)
\eea
Using the appropriate initial conditions for our case,
\be
\delta(R=1)=\delta_0\;, 
\qquad  \frac{d \delta}{dR} (R=1)=
\frac{\dot \delta (t=0)}{\dot R(t=0)} =0 \nonumber\;, 
\ee
it follows that
\[
A_1= 0\;\;,  A_2=-\delta_0/2 \;,
\]
and thus $\delta(R)=f(R) \delta_0$ where 
\be
\label{fullsolution}
f(R) = -\frac{1}{2R^{3/2}} 
\left[ \sqrt{R}(R-3) - 3 \sqrt{1-R} \;\; \mbox{asin}(\sqrt{1-R})
\right] \;.
\ee
The behaviour of Eq.~(\ref{fullsolution}) as $R\to 0$, i.e., close to 
the collapse time, is  
\be
\label{limia1}
f(R) = \frac{3\pi}{4} R^{-3/2}+\mathcal{O}(R^{-1/2})\;.
\ee

\end{document}